# Spiral Edge States in Topological 2D $Sb_2Te_3$ Crystals

T. Carlson[1,2], D. Lazega[1,3], S. McGuire[1,3], M. Medahinne[1,2], Motahhare Mirhosseini[1,2] G. Marcus[1,3], C.C. Dun[1,5], Maria Koleśnik-Gray[4], Vojislav Krstić[1,2,4], D.L. Carroll[1,2*]

[1] Nano/Quantum Technologies Laboratory at Wake Forest University 95 W 32nd Street Winston-Salem NC 27105 USA
[2] Department of Physics, Olin Physical Laboratory, Wake Forest University, Winston Salem NC 27109 USA
[3] Quoherent Inc. 2860 Lyndhurst Avenue, Winston-Salem, NC, 27103, USA
[4] Department of Physics, Friedrich-Alexander-Universität Erlangen-Nürnberg (FAU), Staudtstr. 7, 91058, Erlangen, DE
[5] Lawrence Berkeley National Laboratory, Berkeley, CA, 94720, USA

* Corresponding author Carroldl@wfu.edu

**Abstract**

This work explores the edge-state conduction properties of twinned and spiraled, 2D $Sb_2Te_3$ topological insulator crystals. We show the structure supports spatially continuous edge states within two spiral and knotted edge conduction channels. The channels are connected through a topological defect. This geometry introduces boundary conditions into the 2D crystal that modifies the edge state dispersion. Using a combination of low loss Electron Energy Loss Spectroscopy, Magnetic Force Microscopy and traditional carrier transport measurements we identify a distinct excitation at ~0.155 eV in the spiral that seems to be related to two isolated states within the gap. This state is localized to the crystal edges. The modification of edge state bands using manipulation of a system's quantum geometry may open new avenues for the use of 2D topological systems.

## 1.0 Introduction

The family of layered 2D $M_2X_3$ chalcogenide topological insulators has been of central interest for development of a wide variety of quantum electronics including quantum computing architectures, quantum circuit subcomponents such as interconnects, timers, drivers, and memories, as well as several quantum sensing applications. For example, the use of 2D topological insulators as a qubit platform has been explored theoretically by more than one group,[1] while the use of robust edge states in room temperature quantum circuits has also been demonstrated experimentally.[2] We have recently introduced an analysis of edge state imaging in 2D $Bi_2Te_3$ plates using magnetic force microscopy to examine persistent current flow in the presence of time reversal symmetry breaking. In that work, we have identified a pseudo-Faraday effect in protected edge states, further supporting the application of such states as analogues of classical circuits.[3]

The properties of these 2D chalcogenide materials arise from strong spin-orbit coupling, band inversion near the Dirac points, and time-reversal symmetry. Such characteristics manifest themselves most strongly as surface (edge) states with spin-momentum locking. Such states have a robustness against any perturbation that doesn't break time reversal symmetry (i.e. non-magnetic) and for large, flat 2D plates, they are well behaved anomalous quantum hall systems.

However, edge states are phase coherent macro-states that extend across the object, and so the boundaries, curvature, and fields associated with that object can introduce quantum geometric effects from confinement to phase accumulated localization. A simple example of this is the Corbino geometry applied to a 2D topological insulator. The interaction of the inner edge state and outer edge state dramatically modify the energy levels of the states and open a pseudo-gap at the Dirac point.[4, 5] If we extend this idea further to a Möbius strip even more complexity is added to the electron wavefunction.[6] Thus, an obvious way to extend 2D topological systems into novel circuit analogues is to modify the quantum geometries of their manifolds. And one way to do this is by physical manipulation of the manifold itself.

This is what makes the $Sb_2Te_3$ system so interesting. In nearly every approach to growth of this material, such as chemical synthesis,[7] molecular beam epitaxy (MBE),[8] or MOCVD,[9] for example, this system exhibits a strong tendency to form spiral-morphologies![10] Recent improvements in synthesis[11] have produced highly symmetric crystalline forms in films and in isolated crystals, both of which lack capping layers and exhibit layer-by-layer atomic ordering. We would anticipate that such a structure would yield a pure $\mathbb{Z}_2$ 2D topological class with a complex quantum geometry across its physical manifold. However, it is challenging to anticipate what the overall effect of this geometry might be on the topologically protected edge states in $Sb_2Te_3$, if any. Such states do exist for flat plates of 2D $Sb_2Te_3$,[12] and they exhibit the usual linear dispersion of edge states within its 0.21 eV bandgap in analogy with other 2D topological insulator systems. But relatively little has been done to directly examine the existence and character of such states experimentally for a "topological insulator spiral." We note that there are theoretical works that address spiral boundary conditions in topological-analogous systems, and these give a glimpse of potential state modifications for such geometries.[13]

This work reports on two new observations for the spiral $Sb_2Te_3$ system.

1) Two- and four- probe, electron transport measurements are consistent with metallic conduction channels along the edges of the $Sb_2Te_3$ spiral. Differential conductivity reveals distinct features within the bandgap of the spiral, suggesting isolated bandgap states. Magnetic force microscopy also identifies current induction along the step edges of the spiral terraces that are consistent with continuous current paths formed along dislocations.

2) Low Loss Electron Energy Loss Spectroscopy identifies a variety of low energy features identified as plasmons, polaritons, and excitons variously. However, one loss peak corresponds in energy to features seen in the dI/dV. This feature also correlates well with the known IR absorption feature for $Sb_2Te_3$ spirals at ~ 0.155 eV (1250 $cm^{-1}$).[14]

## 2.0 Experimental

### 2.1 Defining the Crystal

Solution synthesized spiral crystals were used for all the studies of this report. Two distinctly different morphologies were examined: **Figure 1(a)** hexagonal, and **Figure 1(b)** star. These differ only in the size of the basal plane. The specific details of our synthesis route, together with an analysis of properties is given in published reports.[7] **Figures 1 (a, c, d)** show transmission electron microscopy (TEM) images of spiral crystals of $Sb_2Te_3$. **Figure 1(b)** is an SEM image of the spiral crystal.

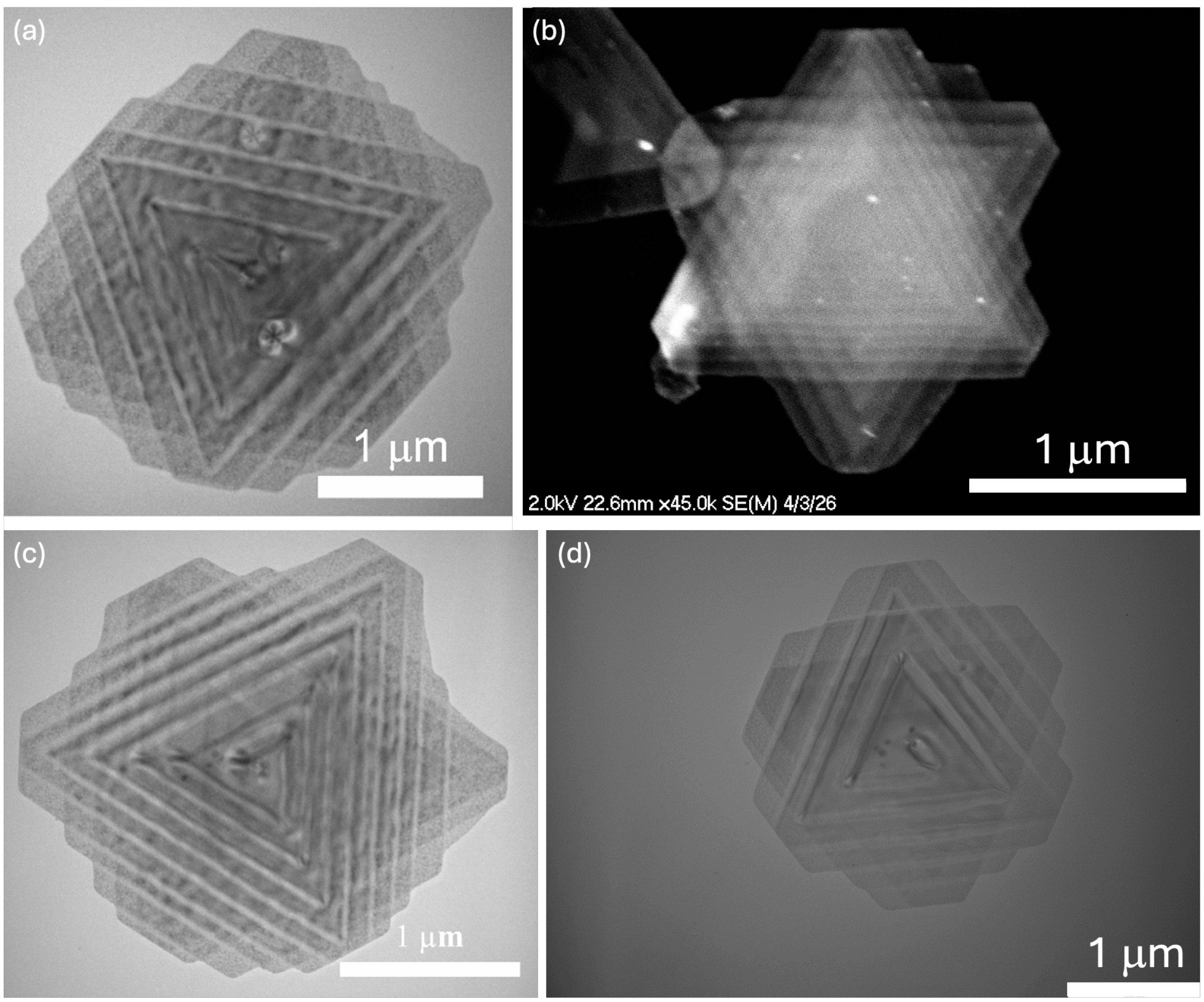

**Figure 1: (a, c, d)** Conventional TEM images of $Sb_2Te_3$ spirals taken with a JEOL 1200EX at 80 keV using an AMT digital imaging system. **(b)** is an SEM taken from a Hitachi 4700S using the secondary electron detector. Note the well-ordered faceting and the variety of the number of spiral loops which occur.

Dun et. al. demonstrated that this twinned crystal grows with a mirror symmetry of the spiral across the habit plane in reference [7]. (**Figure 1**) The steps of the terraces are ~ 1QL (quintuple layer) thick generally. They are connected to a spiral dislocation that runs through the center of the crystal. The spiral edges on top and bottom terminate into the dislocation at the apexes of the crystal. The number of terraces, area and diameter of each terrace, as well as the interactions between terraces can all be controlled through various synthesis modifications. This structure is illustrated in more detail in **Figure 2** and is based upon reference [7].

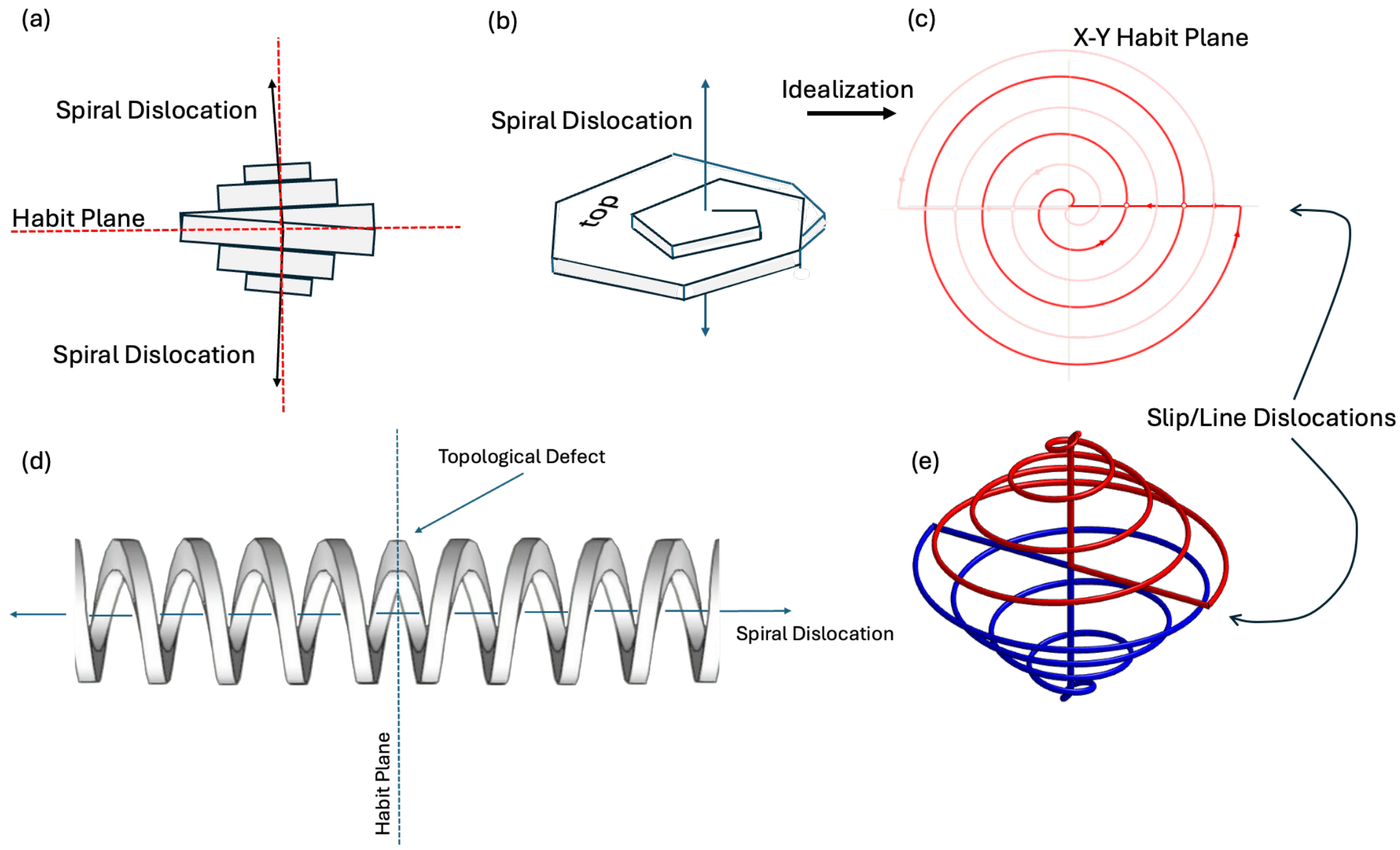

**Figure 2:** Idealized schematics of the spiral crystals of $Sb_2Te_3$.[10] The layers shown in **(a)** and **(b)** are ~ 1nm thick, so a quintuple layer (QL) or the unit cell. **(a)** The crystal structure expresses a strange twinning wherein the upper half of the twin is a mirror image of the lower half. This includes mirroring the spiral dislocation in the center of the crystal, which means the spiral on the bottom half is "other handed" from the top half in its helicity. Looking down from the top **(c)** the edges of the spiral layers can be traced out. Ideally, this edge should support two topological bandgap states each with momentum-spin locking for $\pm \boldsymbol{k}$. Notice however, that these spirals must have a defect where they meet and change handedness as in **(d)**. At this point the crystal itself has a large basal plane extending in XY (with Z along the spiral dislocation) as in **(b)**. In order to support the strain associated with the topological defect at the center of the plate, we anticipate that a Slip or Line Dislocation must be added to the basal planes of the upper and lower halves of the crystal as seen in **(c)** and **(e)**[15] This is consistent with the high resolution electron microscopy of these materials throughout the study.[16] We have idealized this to one Line Dislocation through the center of the crystal intersecting with the Spiral Dislocation.

In **Figure 2**, the overall structure of the system is schematized. Crystals used in this study were grown following the synthesis methods of references.[7] As seen in that reference, not all the resulting crystals will have the symmetries ascribed to our idealized system. Specifically, the intersection of the spiral dislocations with the basal plane dislocations will only be expected for the most symmetric circumstances. It should be noted that for the experiments described herein, we have made every attempt to utilize symmetric and well-formed crystals where the above approximations to structure might hold. And, while a systematic study of crystal *morphology* and its effects on the outcomes of transport, EELS, etc. have not yet been done, we can say anecdotally that badly formed crystals with extreme asymmetries, multiple spiral dislocations, etc. can have quite random properties.

## 2.2 Magnetic Force Imaging

This simple geometric model points to an issue that must be addressed; the lattice dislocations such as the spiral through the center and the slip dislocations along the basal plane, are being treated as though they have 1D topological states associated with them. Specifically, the dislocations have protected conduction channels and they connect the current flow from the spiral outer edges of the crystal into a loop. (**Figure 2 (e)**). Indeed, such a connection is necessary to define the spiral edges as "topological edges" to the structure. The "edges" of the 2D system must enclose the topological insulator fully.[17] We make this dislocation boundary assumption based on theoretical work which suggests that dislocations within the 2D topological insulator (TI) bulk can support a variety of helical and chiral states, modify properties of Hall states, and connect together multiple aspects of topological systems.[18, 19, 20] But, even in light of these references, it isn't clear, *a priori*, that this system should be expected to support the continuity of *circuit-like* states wherein the edge states and dislocation states connect together to form a single conduction channel around the object. This would require that the edge states and dislocation states have similar energies. And even if this does occur, the resulting current flow in the coil could be a challenge to describe. Spin-orbit coupling in a spiral or coiled system may be well studied for spin-spirals [21] but it isn't clear that such systems would be analogous to our topological insulator spiraled crystals where spin-momentum locking exists.

Magnetic force microscopy (MFM) is used in this study to better resolve this issue. Faraday inductance has recently been reported in the conductive edge channels of 2D TI plates.[22] This indicates that simple geometries support states that are analogous to those we suggest in the spiral crystals. Further, in that work, induced current flow is directly related to contrast in magnetic force microscopy (MFM) phase images and this is quantified through a model which comprehensively accounts for damping due to loss mechanisms as well as magnetic scattering within the topologically protected channels. This approach, introduced by our group specifically for TI systems, provides an ideal context in which to test the continuity of the coiled conduction channels of the spiral crystals. If the coil doesn't form a closed loop, there can be no current flow along the edge and thus no magnetic response of the system.

For scanning probe studies, the residue from the synthesis chemistry left on the spirals must be removed, in this case ethylene glycol and polyvinylpyrrolidone (PVP). So, the sample was washed with N-Methyl-2-pyrrolidone (NMP) and ethanol before imaging. Imaging was carried out using two instruments, a Digital Instruments Dimension 3100 (Brucker D3100). The DI uses a line-by-line lift technique to perform lift/phase imaging. In both cases the microscopes use magnetic cantilevers from Nanosensors - PPP-LM-MFMR-10 low moment, coercivity of approximately 250 Oe, remanence magnetization of approximately 150 emu/cm$^3$. The tips were magnetized by applying 4-6 kOe external field parallel to tip axis for 1 minute and removing the field and waiting 5 minutes for stabilization of the tip field.

Imaging of the spirals was carried out on Si/$SiO_2$ (a 100 nm thermally grown oxide layer on doped Si) substrates as well as freshly cleaved HOPG substrates. A comparison of the different substrates, one conductive and one not, shows no difference in contrast formation.

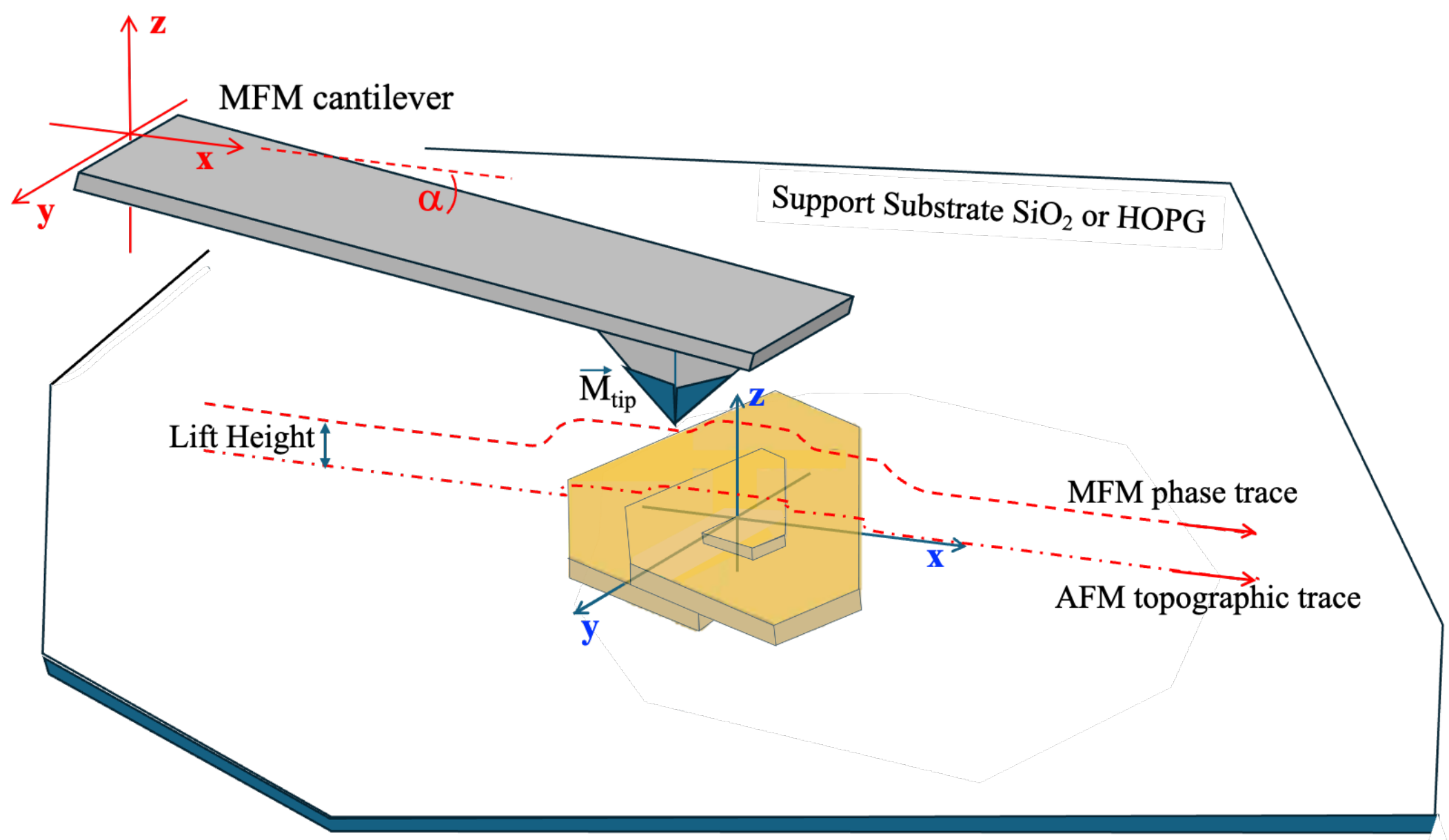

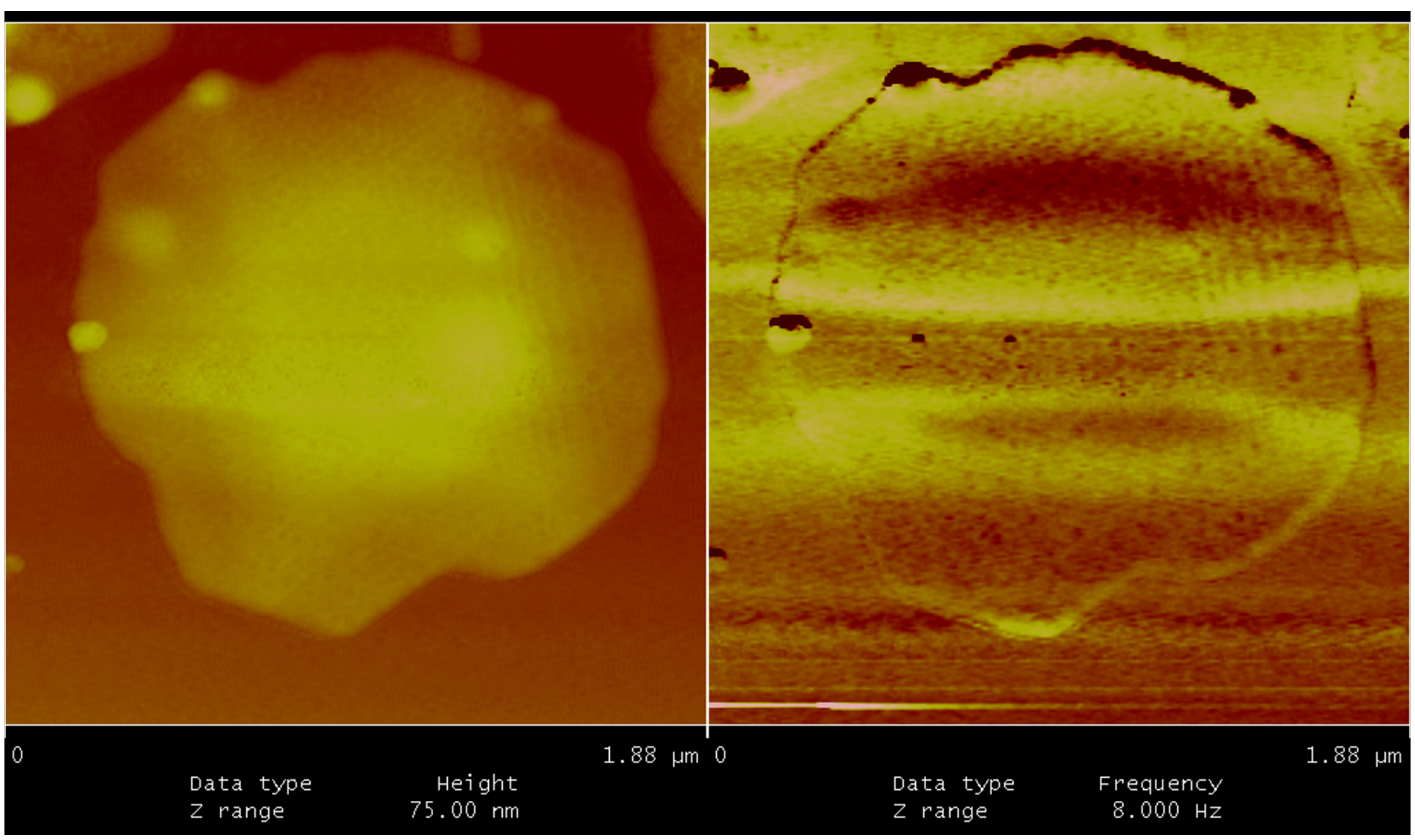

**Figure 3: (top)** Magnetic Force Microscopy (MFM) phase imaging was used to determine the magnetic response of the spiral crystals. AFM and MFM phase images were collected simultaneously using a line-by-line lift mode. First a topographic trace is made using tapping mode AFM, then the cantilever is dithered at some lift height above the surface, and the line trace is repeated using the topography trace to ensure a constant height above the surface **(bottom)** *left* - AFM image of a typical spiral crystal. *right* – phase image showing a magnetic response of the step edges of the crystal. The magnetic edge contrast is caused by faraday-induced currents and suggests that the current flows continuously along each step edge of the spiral.

From reference [22] the phase image derives from two distinct mechanisms which we write in terms of magnetic and electronic components.

$$\Delta\phi = -\frac{Q}{k}\left[\frac{dF_{mag}}{dz_m}\bigg|_{mag} + \frac{dF_{elec}}{dz}\bigg|_{elec}\right]$$

ϕ is the phase between the drive voltage and the response of the cantilever. Δϕ is the phase difference between the phase of the free-oscillation of the cantilever far from the surface, and the phase as measured at a given lift height just above the surface during phase imaging. $k$ is the spring constant of the cantilever, Q is the quality factor of the oscillator, $F_{mag}$ is the force due to the magnetic interactions which derive from induced currents as well as anomalous spin states and localized scattering. $F_{elec}$ are capacitive and dielectric relaxation interactions which are coupled to the metallic AFM/MFM tip.

The magnetic interactions in this case can be quite complex since this is essentially induction into a spiral.[23] However for relatively modest lift heights, we can approximate each spiral plane as a loop of radius r giving:

$$\Delta\phi_{mag} \sim \frac{\mu_0 I_{persistent} m_{eff}}{\pi}\frac{Q_l}{k}\left[\frac{1}{z_m^3}\right] + \int \frac{3\mu_o^2 m_{eff}^2\, Q_l}{4\rho\pi^2 k}\left[\frac{-A\omega r}{z_m^2(z_m^2 + 4r^2)^{5/2}}\right] dr$$

$m_{eff}$ is the effective magnetic dipole moment of the tip, $z_m$ is the height to the tip dipole, ρ is the resistivity, A is the amplitude of tip oscillation, ω is the angular frequency and r is the radius of the loop on the spiral into which the current is being induced. The integral is carried out to the overall radius of the crystal, $R_c$. This assumes a crystal like that of **Figure 3** wherein there are many closely spaced spirals.[24]

$I_{persistent}$ is a "persistent current" of the topological system caused by the tip's magnetic field. However, because the Zeeman energy is very small from an MFM tip this is not the typical persistent current expected of a 2D topological insulator in a magnetic field. Instead, it derives from an intermediate scattering mechanism due to the proximal tip field at the edge of the TI. This means current flow within the system will have some scattering during the imaging process leading to a dissipative term in the Q of the oscillating tip. $Q_l$ is an effective quality factor which includes dissipation losses and damping the Lenz back reaction. A good approximation to the $z_m$ dependence of $Q_l$ is given by in reference [22]:

$$Q_l = z_m^4\, Q \,/\, (\alpha^4 - z_m^4)$$

where α is a constant. As discussed extensively in the references provided, this means that the imaging of currents along the edges of topological insulator plates using MFM is reasonable and that the contrast as we observe here, only occurs when such conducting topological edge states exist.

Moreover, capacitive and dielectric response can be expressed as:

$$\Delta\phi_{elec} \sim \frac{Q_{l-HOPG/SbTe}}{k}\frac{\pi\,\epsilon R\,U_c^{\,2}}{z^2}$$

where R is the radius of the tip, z is the distance between tip and surface of sample, and $Q_l$ is the effective Q over the sample/substrate. $U_c$ is the contact potential between sample and cantilever.

$$Q_l = ½\,kA^2\,/\omega\epsilon\tan(\delta)\,(U_c/z)^2$$

We approximate a $Q_l$ based on dielectric loss.[22] For small sample tip separations:

$$\Delta\phi_{elec} \sim \frac{\pi A^2\,R}{2\,\omega\,\tan(\delta)}$$

So, for the tip imaging relatively close to the sample, the $\Delta\phi$ is relatively insensitive to small changes in the dielectric. Height differences over a step for example will not show up as the strong lines of contrast we observe.

It should be realized that this method of sample preparation allows the $Sb_2Te_3$ to be exposed to atmosphere. O-termination at Te sites on the edges of pores and plates is recognized to develop an interplay between the antiferromagnetic behavior of edge bands and a ferromagnetic phase along those edges In $Bi_2Te_3$.[25] And, this likely to be the case in our samples as well. However, this ferromagnetic phase is only seen for temperatures ~ 1K and even then, it is smaller than the expected fields of induced currents.

It is also important to note that a bandgap of Eg ~ 0.21eV at a temperature of 25 $^{o}$C will ensure the conduction band will have carriers. There isn't a strong magnetic response from across the bulk terraces, suggesting that any induced current there is extremely small. From this we suggest that the contrast features observed on the terrace edges across the spirals as seen in the example of **Figure 3 (bottom-right)**, are associated with current flow along those terrace steps. Given the energies involved, we do not expect charge pumping between the conduction band and edge states. Generally, this should mean some form of continuity along the conduction pathways associated with the terrace step edge and dislocations that might connect them together.

**2.3 Transport measurements**

DC current - voltage measurements were made using standard two- and four- probe techniques. Again, these measurements were made at 293 K meaning that we expect for $Sb_2Te_3$ this should be $1.0 \times 10^4$ to $5.0 \times 10^4$ S/m (100 to 500 S/cm) or an electrical resistivity of ~ 0.2 to 1.0 mΩ·cm, based on bulk determinations.[26] However, this can be strongly dependent on the number of QLs in the structure.[27]

We should say a little more regarding our expectations. Specifically, to account for the layering of the spiral we might approximate the crystal as a stack of plates of decreasing diameter as in **Figure 2 (a)**. We note that for our studies; to make identification of the number of spiral terraces obvious in SEM, we use crystals that have only a few layers on the top and a few on the bottom (generally between 3 and 6). If the interlayer interactions reflect those of simple plain crystals, then this will result in each terrace having a different bandgap based on the *ab initio* calculations of [2727]. This effect is not small with $E_g$ variation from ~ 1eV for 1 QL to 0.21 eV at 6 QL. Thus, even at room temperatures, there is expected to be a potential variation across the different terraces that may cause confinement.

To examine the transport properties of such crystals, the as synthesized spirals suspended in ethanol were drop-cast onto a pre-patterned $Si/SiO_2$ substrate. Again, the spirals were washed with N-Methyl-2-pyrrolidone (NMP) then a thin layer of resist added. Finally, an individual spiral was isolated by imaging through the resist. Traces were added connecting the spiral and the prepatterned contacts at the chip edge using $e^-$ beam lithography on a JEOL IT200 equipped with a Nabity NPGS system. The patterned was developed, metal was added and the resist released.

Care was taken to contact the upper half of the spiral crystal or the lower half with no contact lines cutting across both types of edges as could be determined from the SEM imaging. This was done by making the contacts to the tips of the star structure as seen in **Figure 4 (Right)**. Importantly, not all crystals are clearly resolved and symmetric. For such samples, contacts could not be reliably made and identified to specific positions on the crystal. However, there are a number that are represented by the **Figure 4** geometry. For this study on such examples are used.

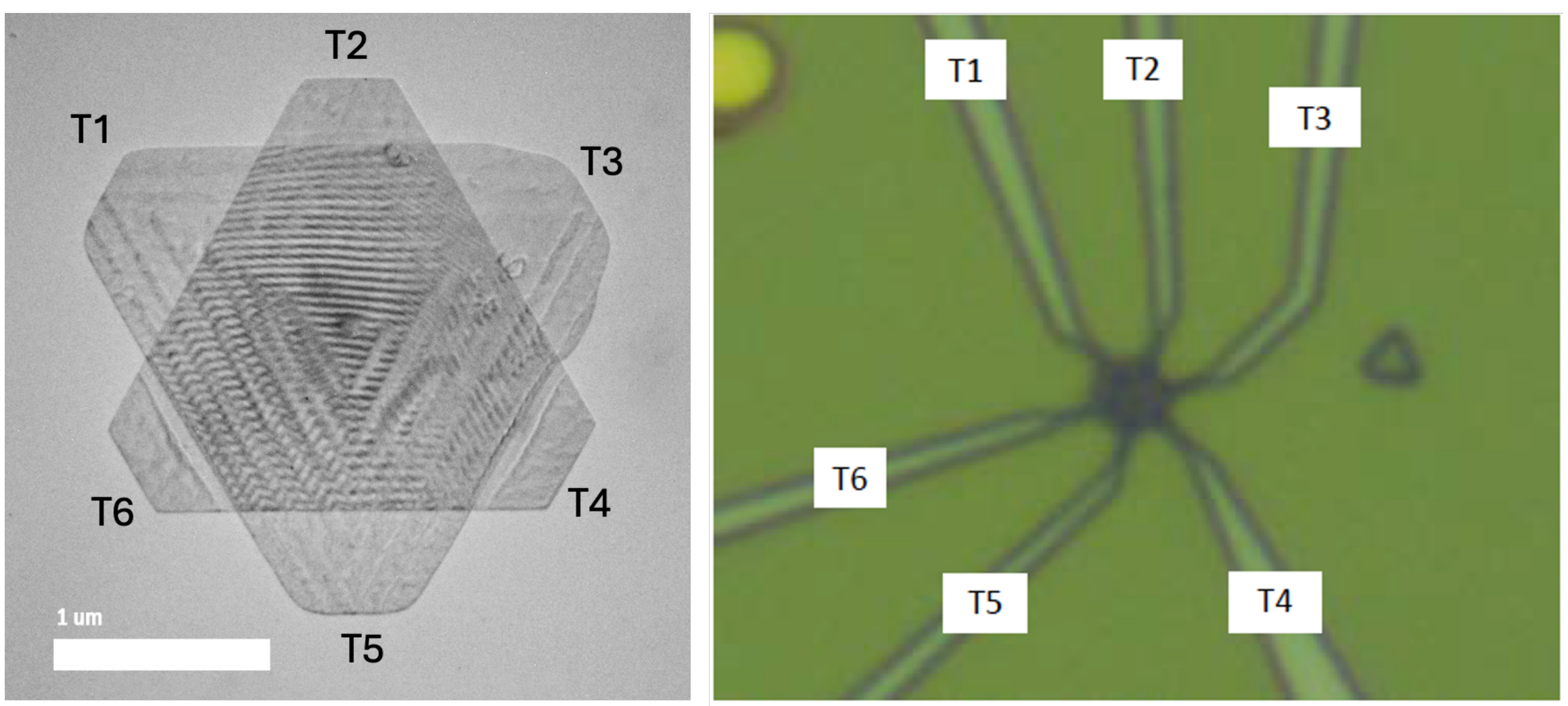

**Figure 4: (Left)** is a TEM micrograph of a typical spiral in which the upper and lower portions of the spiral crystal are easily identifiable. This image shows clearly how the upper and lower halves of the spiral can be identified with specific crystal edge-facets (the points of the star) **(Right)** is the way in which the leads and contacts are added. Care is taken to contact only the very edge of the star tips. Notice that T1, T3, T5 are associated with the top of the crystal. T2, T4, T6, are associated with the bottom of the crystal.

As mentioned, both two-point and four-point probes were used to determine transport properties of the crystals. Note that the exact placement location of the probe/contact onto the crystal will define the current paths that dominate the measured IV characteristics. This is quite clear from **Figure 5** below. Notice that if the probes are places along the upper part of the crystal, the current that is injected will flow along the upper spiral only. There is no applied potential to induce flow across the central defect and onto the bottom half of the crystal. However, if the contacts are made such that one is situated on the top half and the other on the bottom half, the current must flow through the central defect and cross the habit plane following the assumption that the line defects support highly conductive transport channels that dominate transport for low voltages over that of the bulk. For four-point probe measurements we have only three contacts to the top and three to the bottom. There, we will always have current flowing through the center.

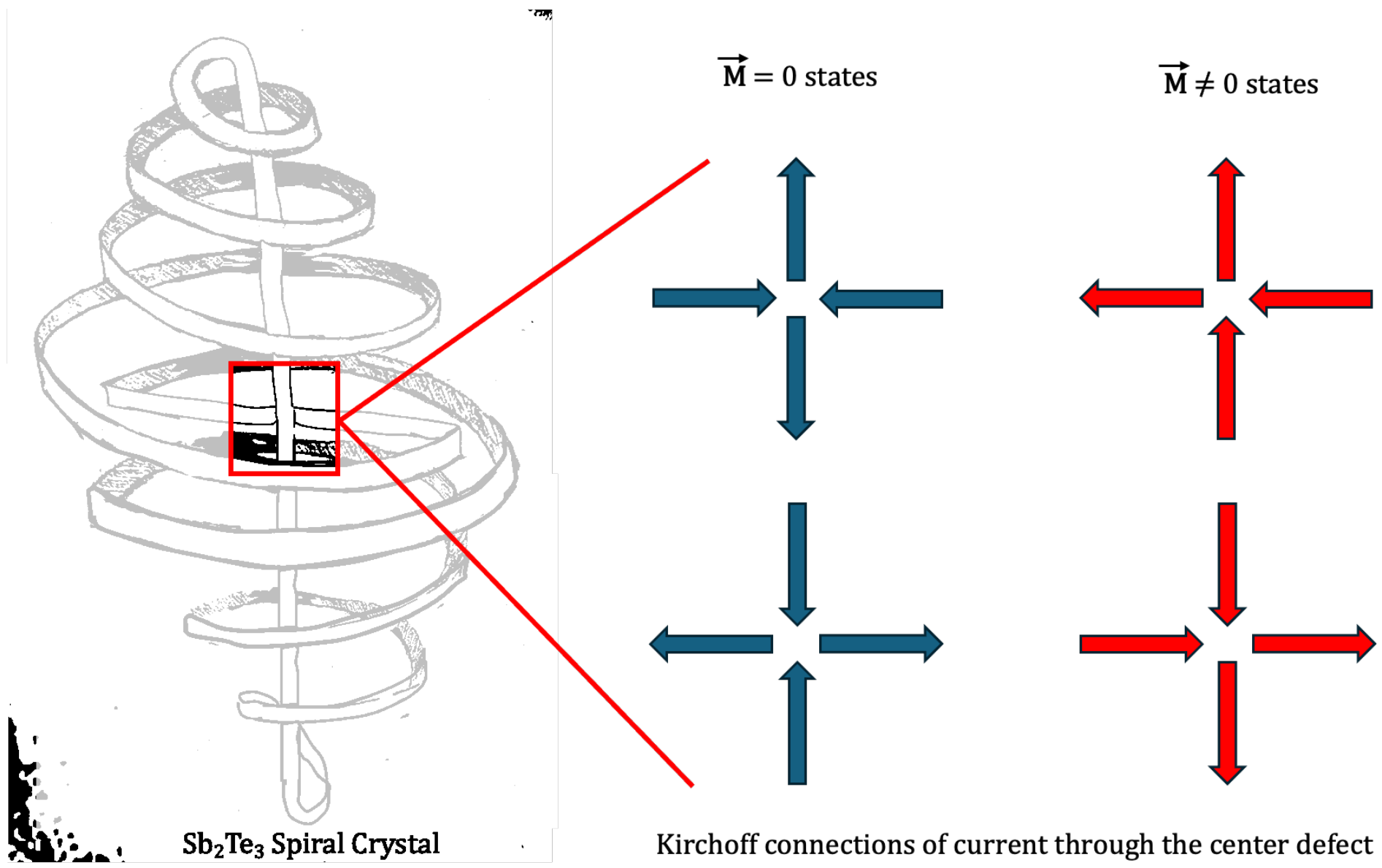

**Figure 5:** There only four patterns of current flow through the system allowed by conservation of charge and by the helicity of the current paths. We note above the patterns (right) that the total magnetic moment of the paths in blue is zero. However, if the current flows in the patterns in red, the object has a finite magnetic moment.

This defines only four possible current patterns relative to the central defect on the spiral dislocation, for low voltages. Referring to **Figure 4** in **Figure 5**, if T1 is used as source and T4 as drain, we expect a current flow like that of the red graphs. Similarly, T2 and T5 as source and drain respectively, also gives the same current flow pattern. That is, current injection from top to bottom of the double coil on T1/T4 or T2/T5 presents current patterns of only one kind (the top red or bottom red depending on the direction of injection). It is interesting that combined spirals of the red flow patterns produce a total non-zero magnetic moment $M_z$. However, the blue current paths allow for induction from the upper coil to the lower, yielding a $M_z \rightarrow 0$ for highly symmetric crystals.

Again, keep in mind this is for low voltage in the IV determination. At higher injection voltages we expect that bulk transport will also occur allowing current to flow along the spiraled planes. For this conduction there will be accompanying nuances such as the possibility of a varying bandgap along the spiral as well as the potential for radial flow from the spiral defect to the edge states.

Because of the complications of four probe placement on such systems, we begin two-point probes to examine current-voltage characteristics, comparing through the center and along the upper planes pathways. This is shown in **Figure 6**. Using the configuration in **Figure 4**, and 60 nm Pd electrodes, the terminal designations (ie. T1T4…) are given. The curves on the left flow the current through the center of the crystal, those on the right, flow current only along the upper edge channels.

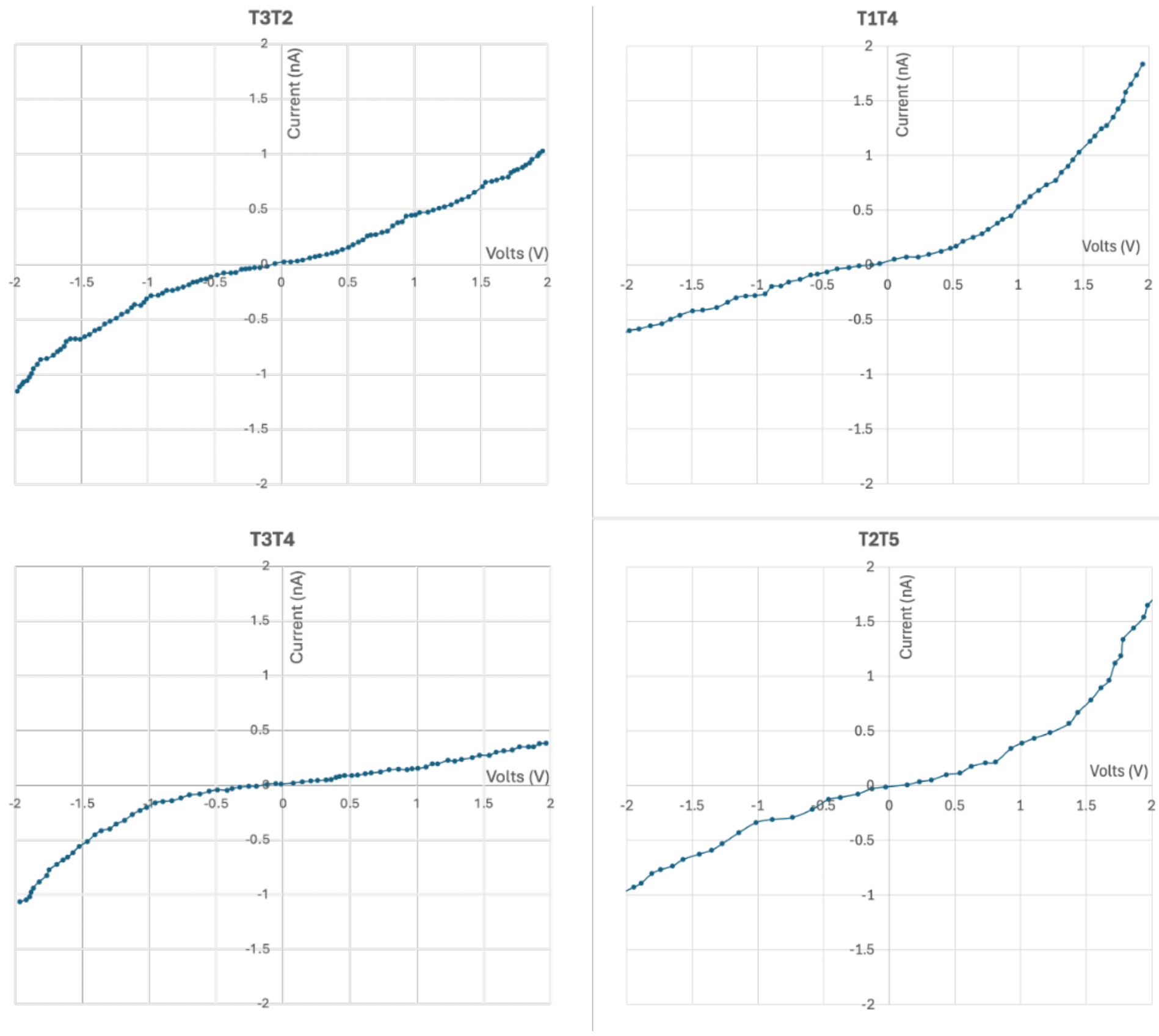

**Figure 6:** Presented here are several two-probe current-voltage (IV) characteristics taken from the configuration of **Figure 4.** T1T4 and T2T5 are contacts to the outer spiral edge on the upper half of the spiral crystal only. Current flows along the edges between these two points. T3T2 and T3T4 connect across the habit plane (from the top of the spiral to the bottom of the spiral). Thus, the current flows along the spiral dislocation through the central defect.

Importantly, two-probe IV curves can give a sense of the interaction of the contact metals with the $Sb_2Te_3$. The bandgap of this material is ~ 0.21 eV, and a close inspection of IV characteristics in **Figure 6** shows a slight asymmetry due to some rectification. We note that increasing voltage ramp speed shows a capacitive effect as well which is not shown here. Within the narrow energy range of -0.5 eV to 0.5 eV, through the bandgap of the material, we observe a smooth and symmetric curve with little trace of an energy gap as expected if the current flows through metallic edge states.

Of the two groups of curves T3T2- T3T4 seem similar and T1T4-T2T5 are similar. But there are differences between the groups. The rectification (asymmetry around zero (0,0)) for the two types of current paths presented is a little different. The T3T2 and T3T4 terminals connect through the central defect along the spiral dislocation. Strong current breakover occurs for these IV between -0.5 V and -1.0 V. For the T1T4 and T2T5 pathways which run along the spiral edge on the upper half of the crystal, notice that the negative voltages are nearly flat. However, for the positive part of the spectra, there is a strong upturn at roughly 0.5 V.

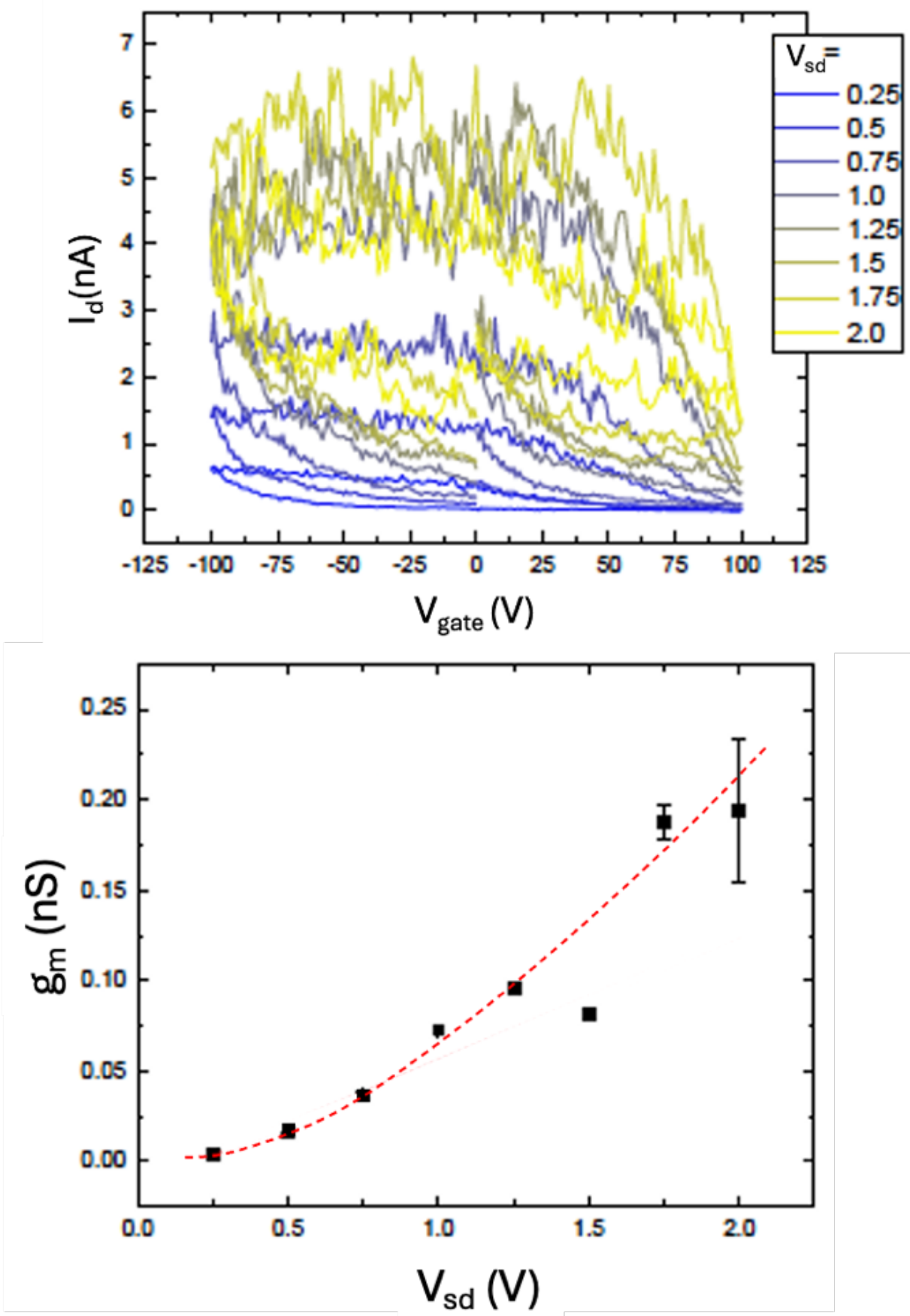

Using the contact configuration described above for a spiral deposited on Si/$SiO_2$ with the addition of a back-gate, we can look at two probe conductivity as a function of applied field on the gate known as transconduction. This allows carrier numbers to be limited. In this case a terminal pairing that connects top to bottom halves of the spiral is chosen. A source to drain voltage ($V_{sd}$) is applied and $I_d$ is measured. For each value of $V_{sd}$ applied, a sweep of the back-gate voltage $V_{gate}$ is made. The raw results are shown in the top panel of **Figure 7**. From this data the ratio of the measured drain current $I_d$ to the input voltage $V_{sd}$ gives the transconduction. Because these experiments use two-point Pd contacts, the relation between $I_d$ and $V_{sd}$ is not linear over the full range of measurement.

**Figure 7:** The transconduction of a spiral for pathways defined through the habit plane. Note the jump in the transconductance curves for gate voltages between 0.75 V and 1.0 V. This also shows that our materials are *p*-type as expected from $Sb_2Te_3$.

However, contact doping, hybridization bands formed with the contact metal, new materials phase formation and defect scattering must all be considered in two-probe transport.[28] So, multiple metal compositions for contacts were tested to better understand the role of issues in the 2D material. Metal compositions[29] used for contacts in these studies were, 40 nm Au with a 10 nm Ti adhesion layer, 60 nm Ni, and 60 nm Pd (above). Each of these present interesting interactions with the topological insulator: the Au/Ti matches the workfunction well according to literature, the Ni is magnetic, and while the Palladium has a larger workfunction difference from $Sb_2Te_3$, it forms a more stable contact.

From bulk literature values: the work function ($\phi$) of $Sb_2Te_3$ typically falls in the range of 5.1 eV to 5.5 eV.[30] $\phi_{Pd}$ ~ between 4.5 eV and 5.8 eV (clean evaporated films like ours ~ 4.95 eV). $\phi_{Ni}$ ~ 5.0 eV to 5.15 eV. $\phi_{Au/Ti}$ ~ ~5.1 eV−5.3 eV (if the Ti is oxidized: ~ 4.3 eV - 4.5 eV which is not the case here).[31] Note that the Au/Ti system in our application is 40 nm of Au and 5 nm of Ti. Thus, barrier heights for all should be very similar.

However, comparative data shows something a little different. In **Figure 8** we compare Au/Ti, Pd, and Ni contacts in the two-point probe configuration. These curves are taken from similar current paths on the crystal.

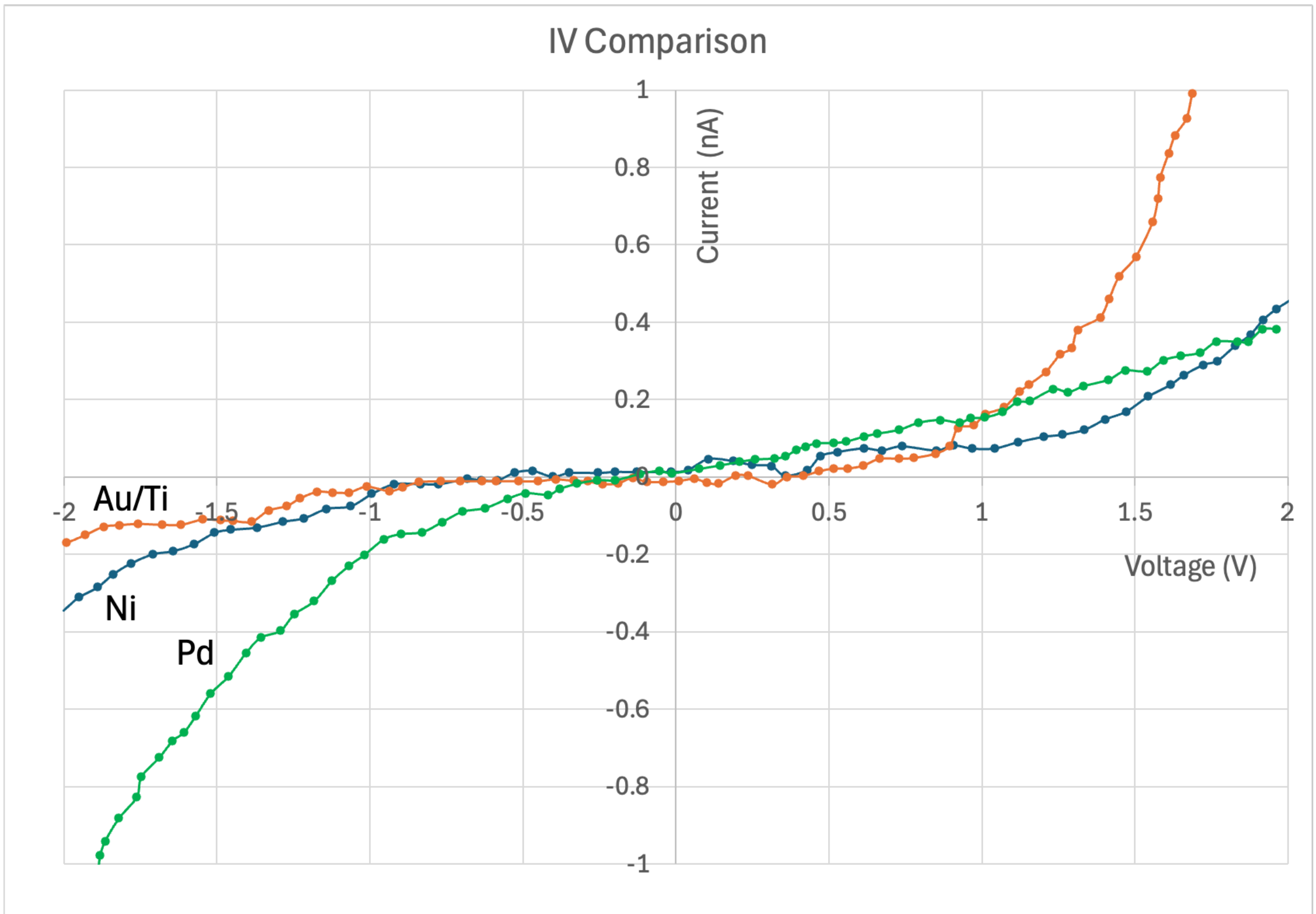

**Figure 8:** two probe IV characteristics of Au/Ti contacts, Ni contacts, and Pd contacts. In all cases the current path is through the center of the crystal.

In **Figure 8** Au/Ti and Pd, both show degree of asymmetry and indicating some asymmetric potential barrier and rectification. The Ni contacted samples, however, are strongly "S curved" and symmetric. The S-curve nature of the IV in these cases is due to band-bending at current injection and while all three express some of this, on Ni shows S-like curves symmetrically around 0V. Notice that Au/Ti is flat on the negative side of V, and Pd is flat on the positive side of V. Normally a linear IV response such as this would be associated with the breakdown of a barrier and injection into a band. But there are two things that stand out in this comparison. The first is that a change in the contact materials changed the carrier being rectified even though the workfunction differences were the same. Secondly, Ni doesn't seem to do this, it has roughly symmetric band bending up to +/- 2V.

It is known that Ti forms phase change alloys with $Sb_2Te_3$. [32] Both Pt and Pd are energetically less likely to form intermetallic compounds with these materials [33,34] Though Pd does interact with $Bi_2Te_3$ [35] and is used in superconducting and thermoelectric applications with Te compounds, relatively less is reported about its interaction as thin films applied to $Sb_2Te_3$. Finally, Ni is well known to diffuse easily into $Sb_2Te_3$[36] and there are reports of room temperature formation of interface intermetallic compound formation for Nickel layers on $Sb_2Te_3$.[37, 38, 39] As with all the metals used here, the morphological aspect of interface formation is ideal for low contact resistance contacts on the substrates.[40] Thus, it is reasonable to suspect that a significant difference in interface formation between Pd, Au/Ti, and Ni, is that Au/Ti and Ni have a tendency toward intermetallic compound formation at the interface with the contact whereas for Pd, this is not energetically favorable at room temperature.

Thus, we can say that the barrier heights of these systems is influenced more by the chemical interactions of the contact metals with the $Sb_2Te_3$ than the workfunction differences. Moreover, it seems that Pd does not dope the $Sb_2Te_3$ system severely and doesn't form parasitic interphase compounds and so for the remainder of the studies we utilize Pd.

Four-probe characteristics of the spiral system using 60 nm thick Pd contacts is shown in **Figure 9**, (relative to T3T2 and T3T4 in **Figure 4**).

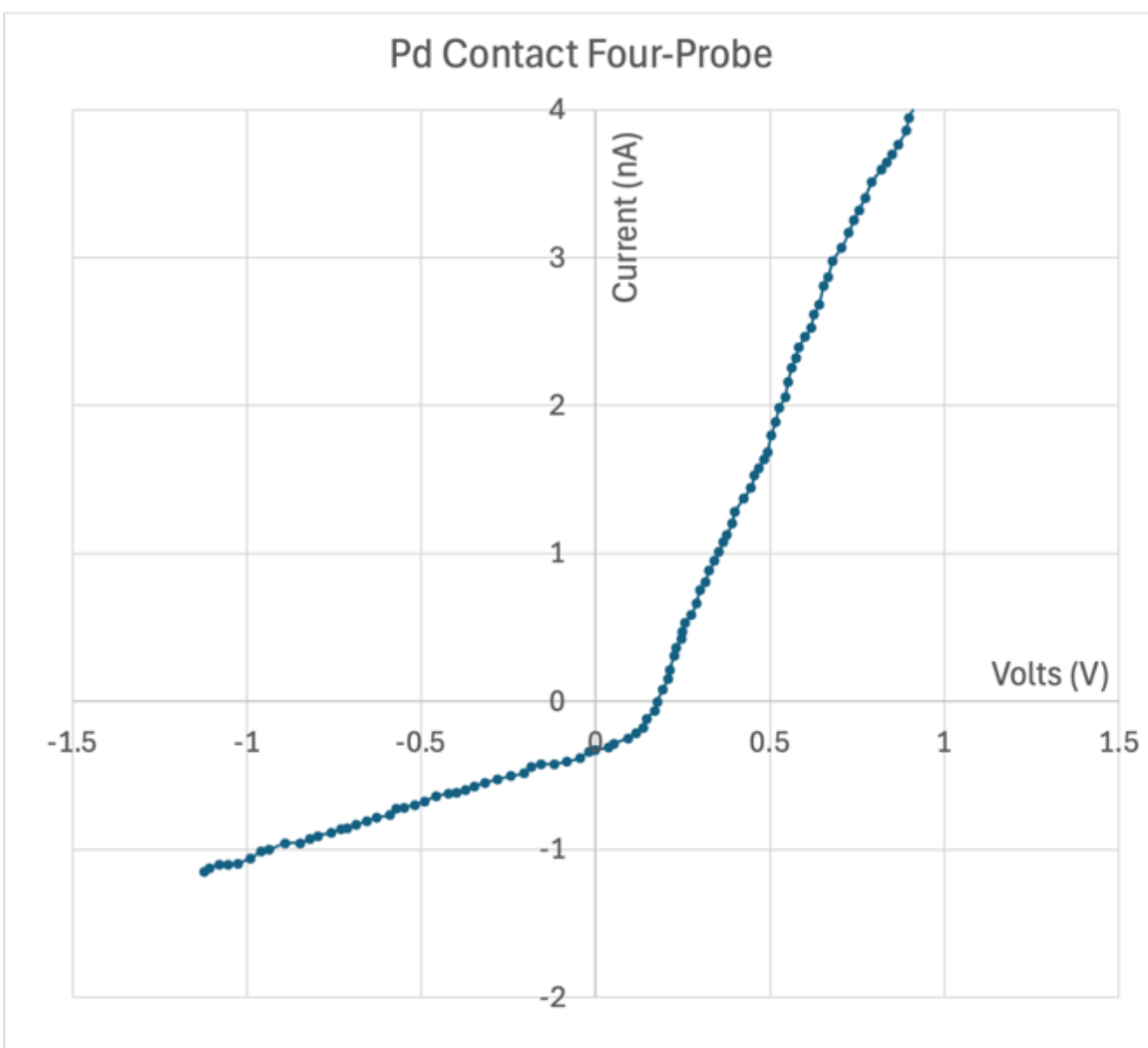

**Figure 9:** Four probe IV characteristics using a Pd contact composition. This was taken such that the source and drain current flows through the center of the crystal, but the voltage is measured across the upper planes. Comparing this with the IV in **Figure 6**, the same shift is seen for measurements made in the same plane, but the overall shape is quite similar as opposed to what is seen for the two- and four- probe measurements in the Au/Ti system. (the shift in I intercept is capacitive)

In the four-point probe method we have little influence from the interaction of the metal with the sample. The scan rate used for this data was fast and so a capacitive offset is noticed. However, more importantly we see a linear response both in the negative and positive part of V. Further, we see two strongly different slopes with a smooth s-curve transition between them within what would be the bulk bandgap of the material. This may suggest for such a system a transition from one conduction regime or conduction channel to another.

A possible interpretation of this is situation is that two different conduction channels are responsible for it. We know that edge conduction in a topologically protected state would be highly conductive and this would be consistent with the positive voltage side of the IV. Band conduction should also certainly occur outside of the narrow bandgap and it should have far more scattering, thereby yielding a lower conductivity as seen in the negative voltages. Near the bandgap there should be band bending as the voltage is applied and this is seen in the s-curve near $V = 0$. So, why would there be a transition?

We can speculate that this comes about because when current flows in the coil, there is a steady magnetic flux through the coils. This means any spin-momentum locked current expected on the edge can flow in only one direction. So, only one direction is allowed for current injection into this high conductivity channel. However, the bulk conduction channels are open to all current directions. Notice too that with two-point probe measurements it tends to be the negative side of the curve that is rectified while the positive side shows linearity. Contact potential barriers then select for the positive current over the negative. This may help explain why we see the high conductivity side always in the positive quadrant.

The four-point-probe IV curves show definite and repeatable features within the envelope of the curve. Taking the derivative with respect to V, (dI/dV) we can get a sense of how the density of states of this low dimensional system might be behaving.

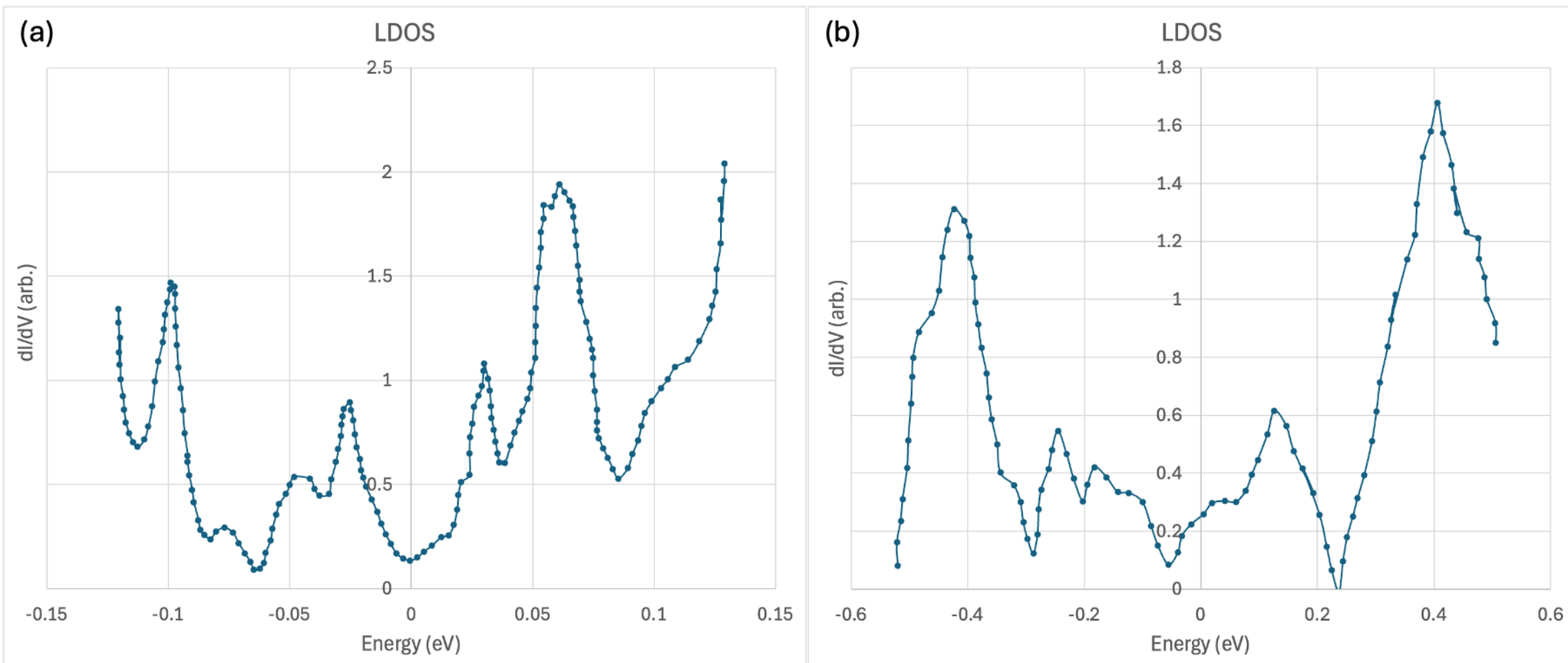

**Figure 10:** Four-point-probe current voltage characteristics can be used to estimate an integrated density of states for the current path under study: dI/dV ~ $DOS_{1D}$. These are differential conductivity curves from two different spirals using the above four-point-probe sample architecture above and Pd contacts.

Using this differential conductivity as a function of voltage as in **Figure 10**, as a proxy for the density of electronic states along the path of the current flow, two different samples are compared. In **Figure 10**, the curve on the right (a) represents a high data density through the bandgap of the material (-0.15 eV to 0.15 eV). (b) is a "zoom out" so that band features can also be seen. The data density in the bandgap is much less so the individual features are not as well resolved. However, in both cases there are distinct bandgap states. For a 2D topological insulator like $Sb_2Te_3$, it is clear to expect a band of energies occurring through the bandgap associated with the edge of the plate. This band of topological should be continuous, but in the DOS above, we clearly see isolated bandgap states. It is also clear that these isolate sharp states cannot be the van Hove singularities typically associated with 1D transport since Umklapp scattering in topological states is forbidden when there is no magnetic field present. Alternative, as we have seen from the references above, dopant states due to the metal contacts can appear in the bandgap. Moreover, it is known[41] that for topological insulators like $Sb_2Te_3$, finite size and other boundary conditions can set conditions on the formation of the edge state bands and/or cause gaps to form in the Dirac cone. Crucially, this is equivalent to saying that the quantum geometry of the object (loosely, the physical space that the wavefunction can occupy) can play a role in setting conditions for the topological states allowed to be supported by that object. Modification of the quantum geometry can then result in modifications to allowed topological states. Such a modification can occur with the addition of local fields, physical restrictions on size and dimension, or spatial orientability of a manifold.

### 2.4 Electron Energy Loss Spectroscopy

$Sb_2Te_3$ spiral crystals were also studied using spatially resolved Electron Energy Loss Spectroscopy (EELS). EELs in Transmission Electron Microscopy (TEM) is becoming a widely accepted tool for the study of 2D chalcogenide topological insulators. It is powerful due to its ability to probe at the nanoscale, electronic and optical properties of these materials providing insight into collective excitations (plasmons), interlayer hybridization, and topological edge states inaccessible by bulk optical methods.[42]

The system used for mapping valence EELS was a ThermoFisher Titan 80-300 aberration corrected scanning transmission electron microscope (STEM) with an energy monochromator and a Gatan spectrometer, capable of point-by-point full spectrum imaging with an energy resolution of 0.15 eV. The convergence semi-angle was 19.3 mrad and the collection semi-angle was 71 mrad. The spirals were deposited from solution and imaged using Carbon TEM grids (Ted Pella Inc.) with primary beam energy of 80 keV. The 0 to 10 eV valence loss spectrum is shown in **Figure 11**. Subtraction of the Zero Loss Peak (ZLP) was done using an exponential fit subtraction as outlined by Hatchel et. al. for this energy loss range.[43]

There are several excitation features, the highest energies are known. However, three excitation features resolved at the lower energies of the window, are of interest in this study. They are marked f1 in the 0 -1 eV window, f2 in the 1 - 2 eV window, and finally f3 in the 2 - 4.5 eV window, in **Figure 11 (a)**. Note that **Figure 11 (a)** shows both the raw data set, before ZLP peak removal, (shaded lighter) and the spectra with ZLP removed (same colors but darker shaded). These spectra were collected with the electron beam positioned at different positions relative to the $Sb_2Te_3$ spiral. Specifically, the spiral was imaged across the edge of basal plane terrace and spectra were taken for each pixel. This way the spatial distribution of EELS characteristics can be determined.

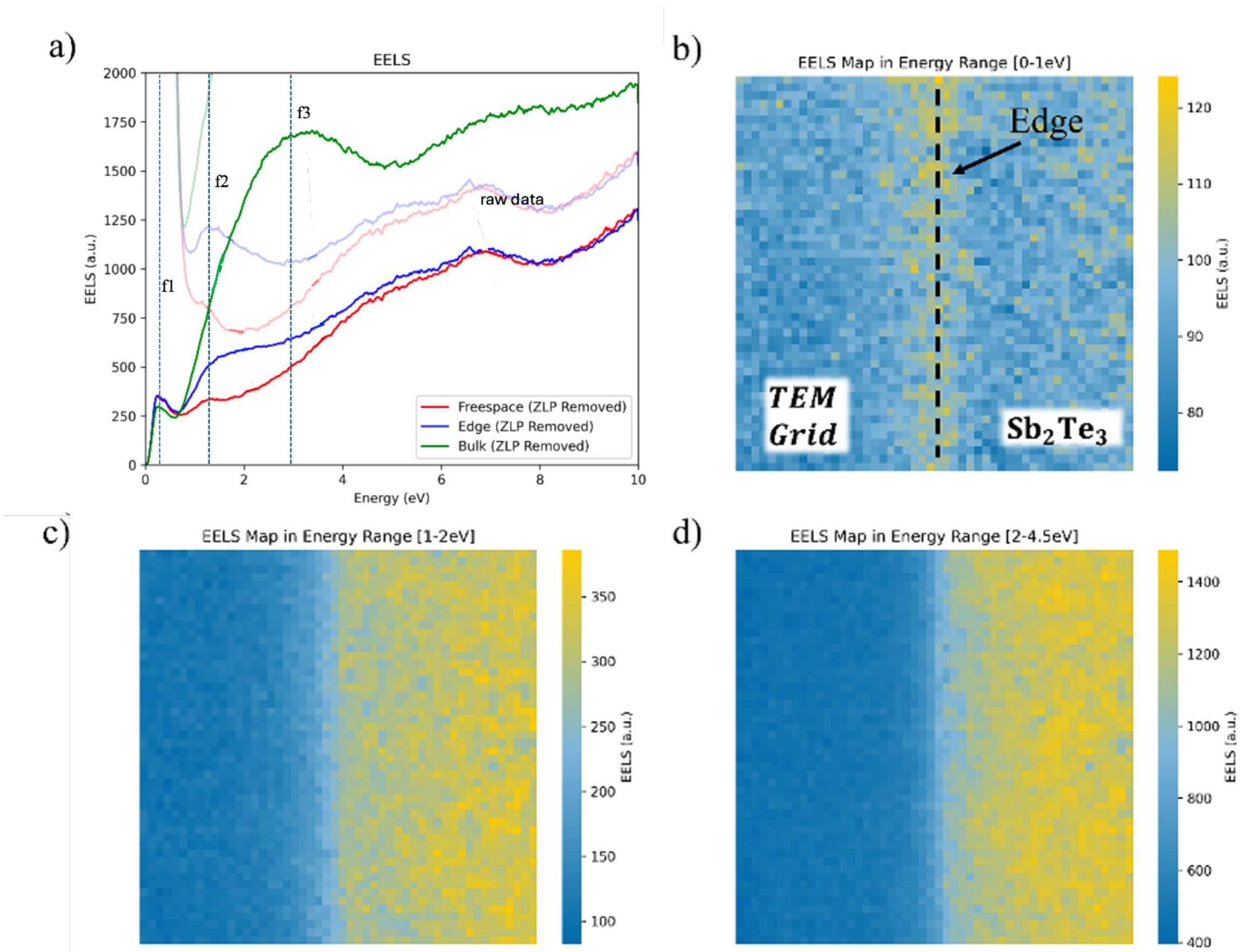

**Figure 11:** Valence EELS on $Sb_2Te_3$ spiral edge. (a) EELS spectrum taken of $Sb_2Te_3$ spiral edge (blue), bulk (green), and in free space (red). (b) EELS map of $Sb_2Te_3$ edge with pixel contrast set between (0-1 eV). (c) EELS map of $Sb_2Te_3$ edge with pixel contrast set between (1-2 eV). (d) EELS map of $Sb_2Te_3$ edge with pixel contrast set between (2-4.5 eV).

Referring to **Figure 11 (a),** the features at ~ 6 eV and its low energy shoulder ~ 4 eV (not marked) are seen in spectra from the bulk, the edge and just outside of the $Sb_2Te_3$ crystal. These are well known plasmon modes and were predicted theoretically by I. A. Nechaev *et. al.* in 2015,[44] as were the features at ~10 eV and ~15eV. These plasmon modes are associated with bulk band transitions, assignments that are now widely accepted.

The feature f3 at ~3 eV is broad and is likely composed of the lower energy f2 and the shoulder seen at ~ 4eV. This leads to an apparent shift in the feature and this is consistent with the assignments of other researchers.[45] f2 (~1 eV) is associated with the edge of the crystal. Its strength, relative to the 4 eV bulk feature, falls off as spectra are taken "deeper" onto the plate and away from the edge. It can also be excited with the beam just outside of the edge of the crystal, which is not unusual for charged or polarized excitations. Moradifar et. al. identify this low energy f2 as an edge plasmon[45] possibly with energy modulation[46] due to the strange morphology of the spiral and its stacking of levels.

**Figure 11 (b) (c) (d)** use an energy filter to create an image of the excitations. (b) is an EELS map of the electron loss in the 0 eV to 1 eV window. The map is taken across the edge of the crystal. The faint contrast along the edge is due to the f2 feature. (c) shows the EELS map of the 1 eV to 2 eV window which captures some of the f2 and the overlap from the f3 feature. (d) fully captures the bulk plasmon and filters out the f2 feature. Notice the very subtle reduction in contrast at the edge from (c) to (d).

The localization of plasmon modes along the edge of the system seems to support the idea that metallic states exist at the edges of the crystal. We might therefore tentatively suggest it as evidence of a topological conduction channel. And the excitation energy is consistent with a low dimensional plasmon calculations in such states for similar topological systems.[47, 48]

f1 is very near the ZLP and near the restrictions of resolution limits. Thus, low loss EELs was investigated using the Nion Hermes microscope at ORNL. This is an aberration-corrected (MAC)-STEM/EELS, equipped with a Cold FEG and ultrahigh vacuum column, a next generation $\alpha$-monochromator and 5th order aberration-corrector. The system was operated at 80 keV. Quantifoil TEM grids were used for the heightened transparency at the energy levels of interest. Since Quantifoil is a glassy carbon film, we expect a relatively broad smooth EELS response in the energy range below 1 eV[49] However, we do note that there are not many references on glassy carbon in this energy loss range.

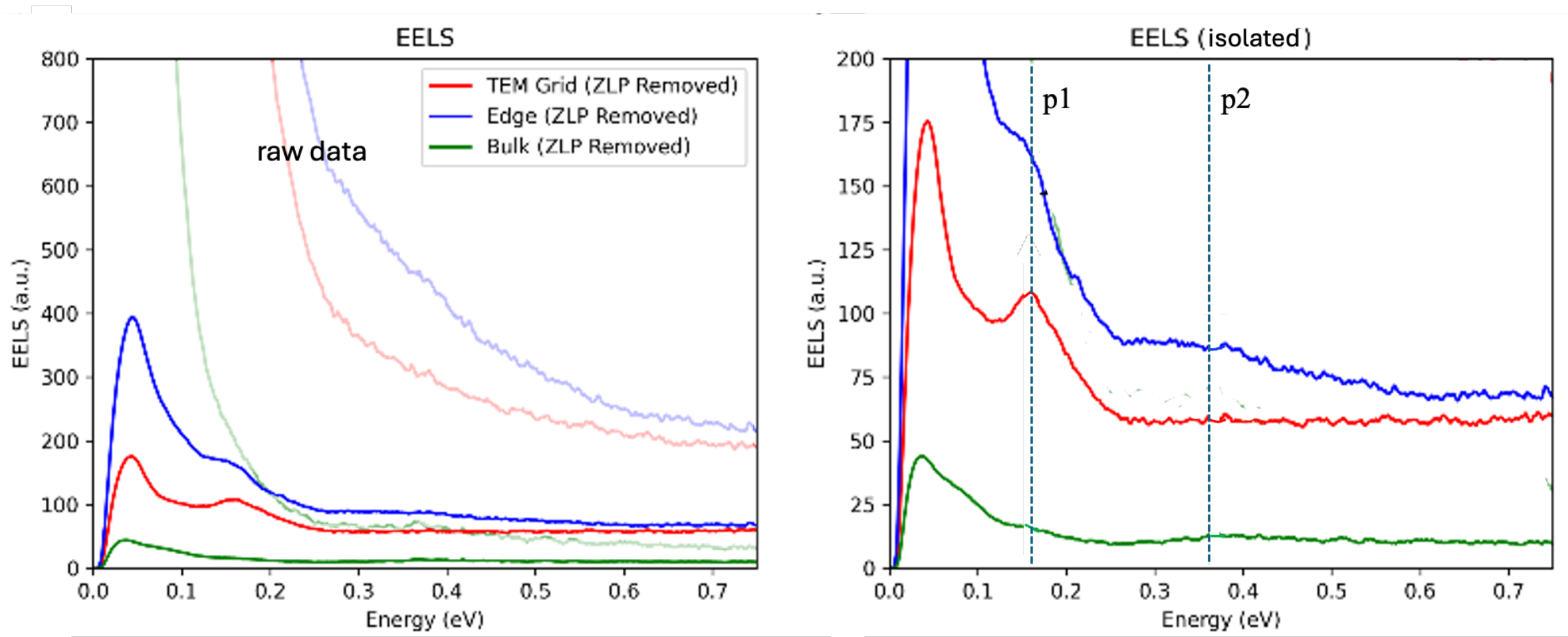

**Figure 12 (left)** the low loss EELS spectra for three different positions on the spiral together with the raw data so that the ZLP removal effects can be seen. **(right)** an enlarged image of just the EELS peaks with the removed ZLP. There are two features, one within the bandgap of the material in energy, the other just outside of this energy window: p1 and p2. The green spectrum represents a signal from the bulk of the material. The blue is a spectrum that contains mainly the edge of the material plate, and the red is taken just off the edge of the material and includes some contribution from the TEM grid which we take to be diffusive.

Two prominent features are observed in the energy window from the ZLP to 0.5 eV. The first ~ 0.16 eV is labelled p1 the second occurs at ~ 0.38 eV and is labelled p2. There may be a small unresolved feature or shoulder at ~0.1 eV, which is close to a known surface plasmon polariton of $Sb_2Te_3$ (0.118 eV).[50] Each of these energies occur within a range than might be expected for phonons in a 2D system such as $Sb_2Te_3$.[51] $Sb_2Te_3$ has the known phonon modes:[52]

- $E_g^1$ (In-plane): 40.7 $cm^{-1}$ (1.22 THz) 5.0 meV
- $A_{1g}^1$(Out-of-plane): 68.0 $cm^{-1}$ (2.04 THz) 8.4 meV
- $E_g^2$ (In-plane): 113.7 $cm^{-1}$ (3.41 THz) 14.1 meV
- $A_{1g}^2$ (Out-of-plane): 166.3 $cm^{-1}$ (4.99 THz) 20.6 meV

There will be some slight shift in energies and the broadening due to confinement induced phonon softening or hardening which is expected with the nanoscale features of our materials.[53] These are the Raman active phonon modes, with other non-active branches being lower in energy generally. There are no Raman active lower energy phonon modes in the bulk of the material. So, it would be hard to identify p1 or p2 with pure phonons.

We note that the strength of the 0.38 eV peak appears lower in the bulk (green curve) than it on the edge of the crystal. This is due to thickness effects as discussed in the references.[51,52] When this is taken into account, the overall ratio of the p2 peak to others is stronger in the bulk than off-edge or at the edge. This suggests it is related to some bulk transition or excitation. However, this energy of 0.38 eV – 0.4 eV is a difficult energy to account for. It doesn't seem to match any direct transition, any plasmon, phonon, or polariton for the $Sb_2Te_3$ system. But it does come close to predicted energies of a Wannier exciton in $Bi_2Se_3$[54] and other chalcogenides.[55] This is further reflected in the excitonic quasi-particle gap of $Sb_2Te_3$ / $Si_2Te_3$ heterostructures.[56] Thus, it seems reasonable to consider a surface excitonic state for the p2 feature.

From **Figure 12** we suggest that the 0.16 eV peak is strongly confined to the edge of the spiral. In comparing the excitation of this feature on the edge (blue) and just off the edge (red), we note that off the edge the phonon peak (p2) falls away but p1 is still excited. This feature becomes significantly stronger as the beam is placed on the edge of the plate and the phonon p2 is also excited as expected. But as we probe deeper onto the plate and away from the edge, notice that p1 falls off relative to p2 suggesting a localization effect.

This can be seen in **Figure 13** as well, which spatially maps the positions where the spectra of **Figure 12** originate. In the top panel of **Figure 13**, the faceted edges of the spiraled terraces are clear. The circular feature that surrounds the image is the Quantifoil glassy carbon TEM grid. Note that the "bulk" collection area (shown in the right-most panel below) actually probes multiple layers and so thickness effects here will be significantly different to that seen on the edge which is going through only 1 QL of the spiral sample. This is seen in the central lower panel. Finally, the off-edge area does include some sampling from the glassy carbon but does not intersect the spiral. As discussed extensively in reference [46], near-field excitation of EELS features can be enhanced by a number of mechanisms when multiple field concentrators are nearby.

We note also that the map-images shown in **Figure 13 (bottom)** are taken using an imaging filter such that only electrons in the 0.1 eV to 0.5 eV window are allowed through. The ZLP has been removed from the spectra but not from the imaging so generally, they do not account for the loss of beam signal due to multiple scattering etc. as the beam scans over thicker areas of the material. This can already be seen in the spectra of **Figure 12 (left)** where the raw data shows a reduction in the total electron flux through the bulk of the material.

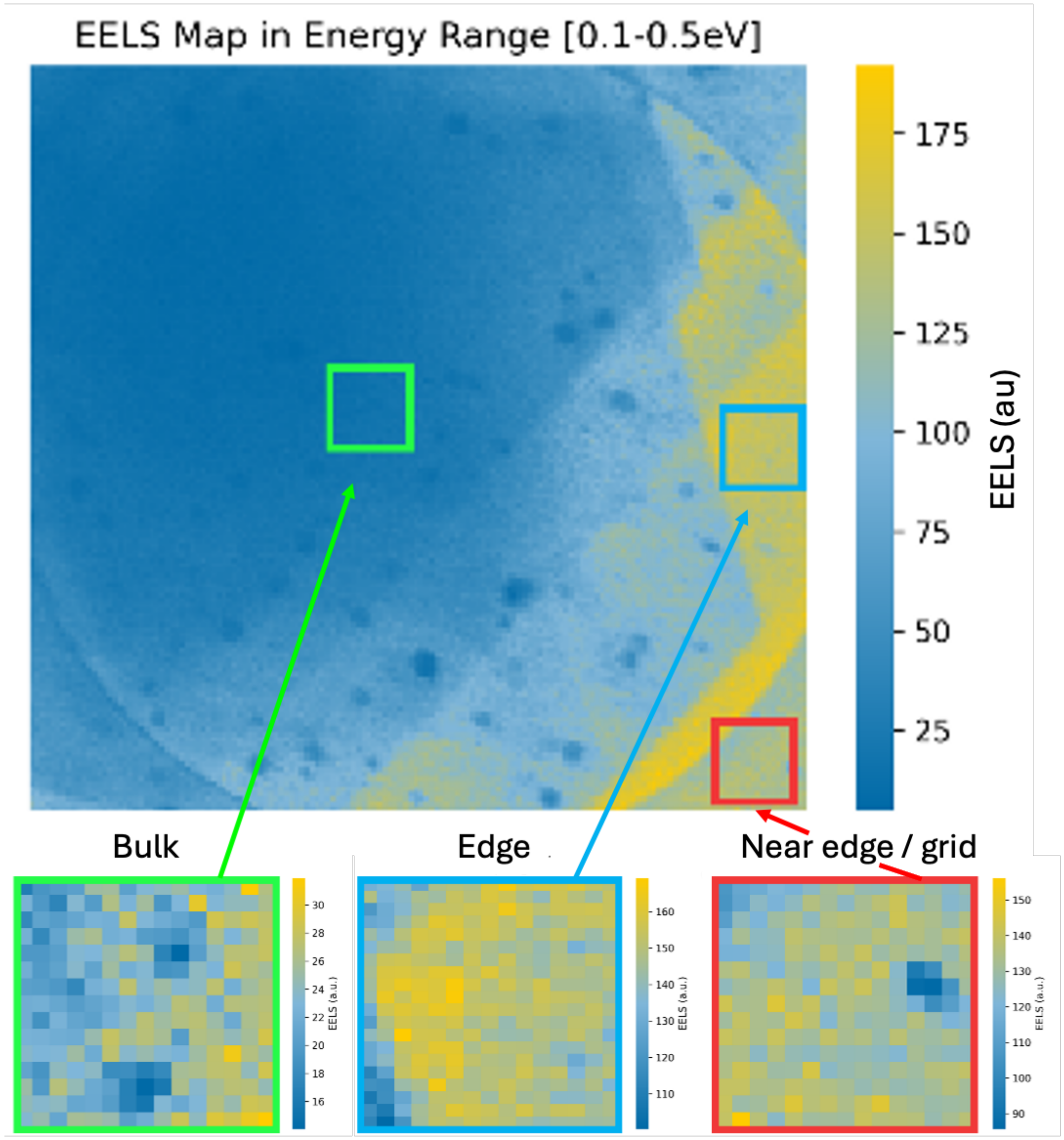

**Figure 13:** a spatial map of the locations from which spectra were collected. These correspond to the spectra in **Figure 12** and the images are energy filtered for the energy range of 0.1 eV to 0.5 eV loss, so they include contributions from p1 and p2.

**Figure 14** examines more closely, the distribution of the p1 feature using higher spatial resolution. p1 is generally stronger at the edge of the $Sb_2Te_3$ than in the bulk or near-edge. These spectra have not been corrected for thickness effects, however, it is clear that the p1 peak gets stronger as the edge is approached from the bulk. As spectra are collected off-edge the signal weakens again. So, at the TEM grid the p1 feature is still stronger than it is in the bulk of the crystal, but the electron flux is also quite a bit stronger there due to the differences in thickness.

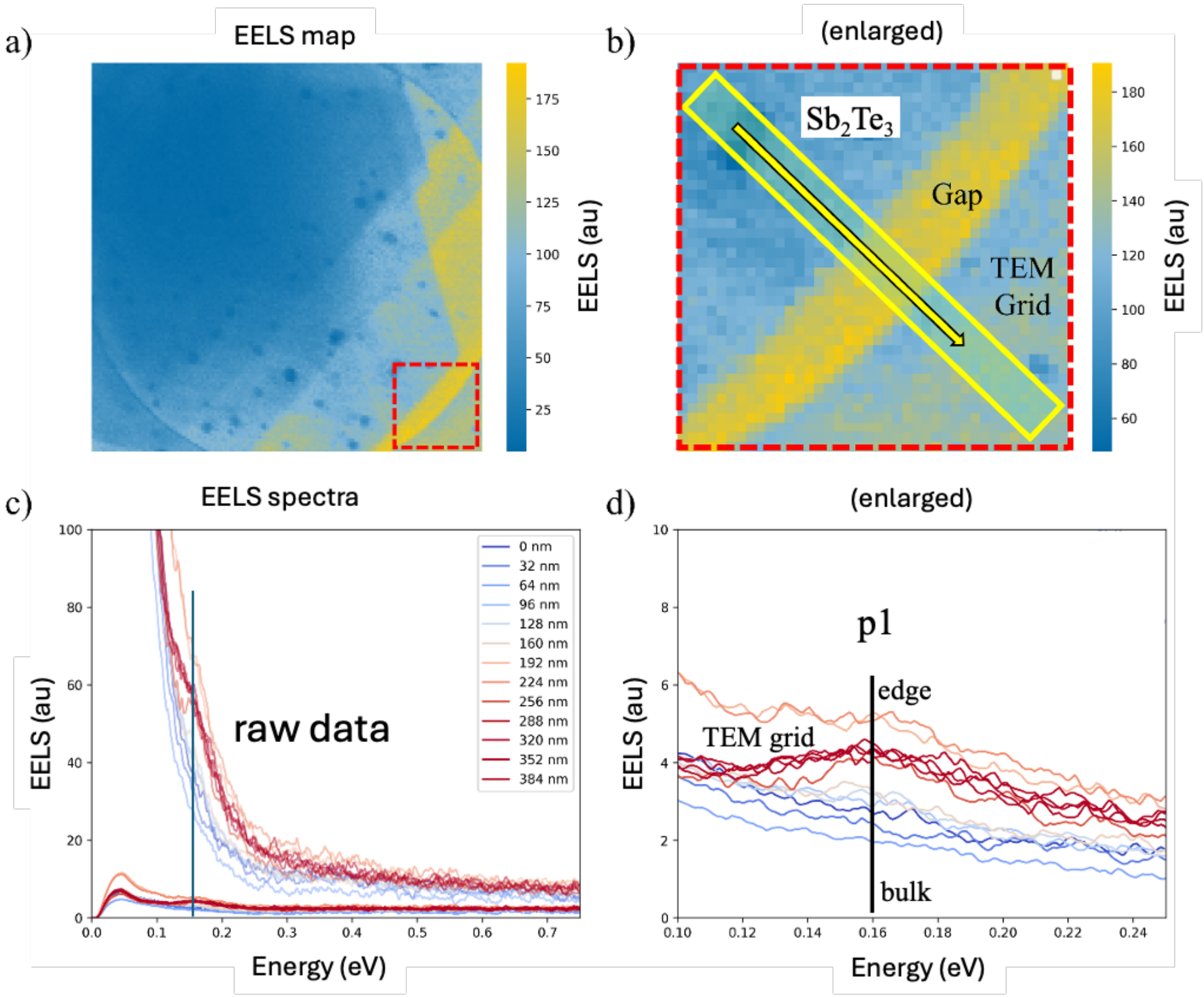

**Figure 14:** Isolating the low-loss EELS p1 feature. (a) the EELS map of the $Sb_2Te_3$ spiral edge. (b) a close-up view of EELS at the $Sb_2Te_3$ edge. Spectra are extracted along the direction of the arrow in the box shown. (c) the EELS spectra taken from $Sb_2Te_3$ bulk, to the gap, and then the TEM grid. Dark blue indicated data taken from $Sb_2Te_3$ bulk with colors approaching red representing displacements approaching TEM grid. Both the raw data and the data with the ZLP removed are shown. (d) and enlarged section of the EELS spectra with p1 isolated.

What we can say about p1 is that it is localized. It is associated with the edges of the spiral. Its energy is inside that of the bandgap of $Sb_2Te_3$. And this energy corresponds rather well to the 1250 $cm^{-1}$ (~ 0.155 eV) resonance seen in IR absorption of spiral crystals of $Sb_2Te_3$ and associated with the Drude plasma edge.[1414] However, as noted in reference [14] this is a rather sharp resonance, and in the EELS spectra we see it manifested as a loss peak unlike what is seen for direct bandgap transitions.[57] We suggest that this can reasonably be interpreted as a transition between isolated or quantized states within the gap and not a transition into the conduction band of the material.

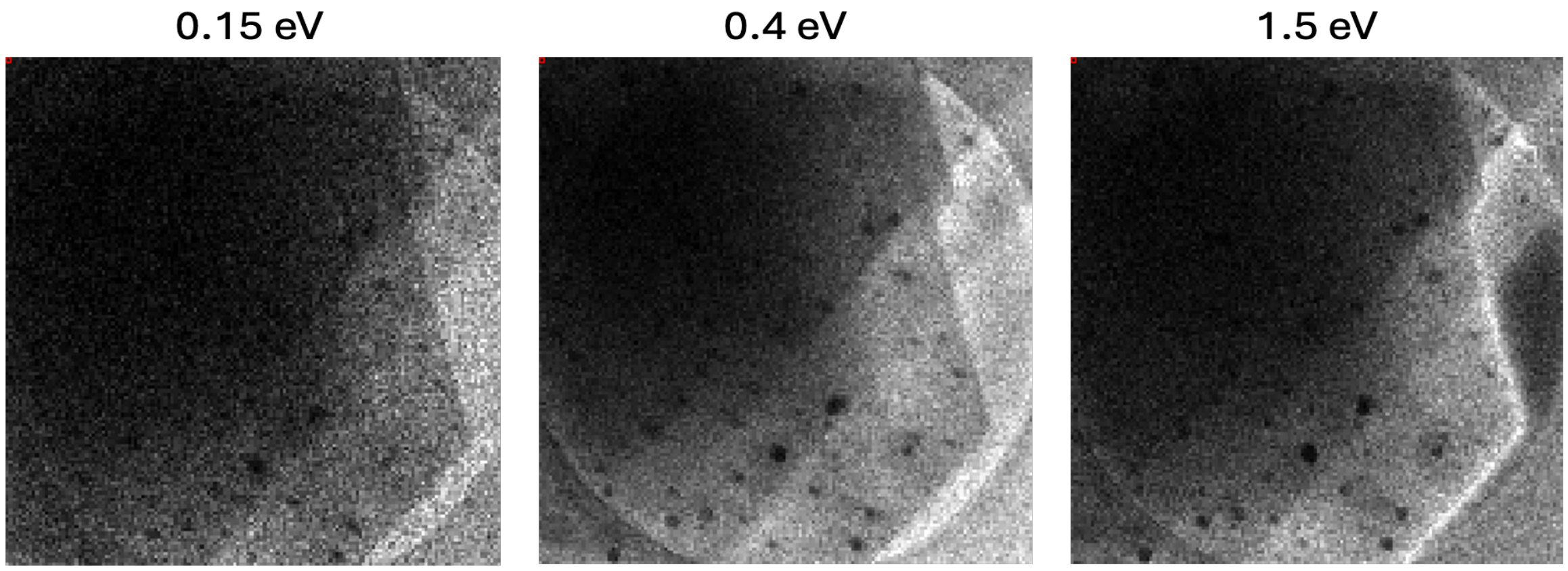

**Figure 15:** These energy filtered images show the spatial distributions of the excitations we have identified as: inter-bandgap transition (~0.15(5) eV), excitonic (~0.4), and surface plasmon (1.5 eV) - captioned above each frame.

**Figure 15** is energy filtered imaging covering three specific energy ranges: ~ 0.15(5) eV, ~ 0.4 eV, and ~ 1.5 eV. Each of these are identified above. The plasmon (1.5 eV) is the strongest contrast and clearly associated with the edge of the spiral. However, the 0.15 eV state is weak and edge contrast can barely be seen. Similarly, the very slight brightening of the plane in the middle panel (0.4 eV) shows the association of this feature with the bulk.

### 3.0 Discussion and Conclusions

We might expect the $Sb_2Te_3$ spiral to behave analogously to previous work with 2D TIs which exhibit quantum spin-Hall behavior in accordance with the BHZ model,[22] expressing quantized spin Hall conductivity and spin component $S_z$ conserved. This investigation has examined some of the basic electronic properties of 2D topological $Sb_2Te_3$ spirals using the same approaches as earlier work on 2D plates. It suggests that the complexity crystal geometries might modify properties that emerge from the band topology of a system. Yet, transport, EELS, and MFM measurements were all done at room temperature which suggests that the stability of such states remains intact and aligned expectations from topological systems.

From electron transport measurements the differential conductivity of **Figure 10** gives a strong suggestion that a pair of isolated doublet states exist within the bandgap. They are positioned at roughly: (-0.05, -0.025) eV and (0.06, 0.03) eV. The upturns at ~ ±0.1 eV mark the band edges. These are, of course, rough estimates because these dI/dV curves have doping due to metal contacts convoluted into them. However, the high energy parts of these doublets come close to the observed 1250 $cm^{-1}$ feature of IR absorption[30,31] and our observed 0.155 eV in EELS feature. So, while the metallic-like conduction channels on the $Sb_2Te_3$ spirals, is consistent with an extension of the BHZ picture[58] from simple geometries, the spiral seems to introduce selection rules into these bandgap states.

To understand this, we introduce a simple scheme in **Figure 2**, comprised of two edge-spirals (an upper and lower) connected together through dislocations in the habit plane and a central spiral dislocation through the center of the crystal. The connectedness of current paths across the crystal has been examined using MFM techniques to spatially map the distribution of magnetic fields. As we have presented it, we suggest that this model system results in quantum-Kirchoff selection rules for the current flow. We hypothesize that this selectivity together with proximal hybridization of edge states along the terrace steps [22] yields an isolated two-state edge conduction characteristic through the formation of a pseudo gap in the edge state dispersion analogous to those predicted theoretically.[59] If this is the case, then the double spiral topological insulator system has an unusual and speculative property: one current solution through the system yields no magnetic moment and spin orientations in the opposing loops add together to give a net $S_z = 0$. However, in the alternative current configuration spin is maintained across the topological defect, and the object also has a magnetic moment for that spin to interact with. It is a spin ½ anomalous hall object.

These results may present an important direction toward a room-temperature, topological insulator quantum circuit architectures. More specifically we suggest that the interplay between quantum geometries and band topologies in BHZ materials can yield an important tool in tailoring stable and robust quantum states in high temperature systems.

**Authors Contributions**

D.L. Carroll, Experimental design, data analysis, manuscript preparation
T. Carlson, EELS and transport data collection, data reduction,
D. Lazega MFM imaging, data reduction
S. McGuire, MFM imaging, materials synthesis, data reduction
M. Medahinne, data reduction and data analysis
Motahhare Mirhosseini, Data reduction and calibration of MFM
G. Marcus, materials characterization and materials synthesis
C.C. Dun, guidance in synthesis, data reduction, manuscript preparation, support data
Maria Koleśnik-Gray, Transport data collection, data reduction
Vojislav Krstić, Transport data reduction experimental design, manuscript preparation

**Data availability**
Data is available upon reasonable request.

**Conflicts of Interest**
This work was funded in part by Quoherent Inc. Quoherent is a startup company in Winston Salem NC and a corporate spinout of Prof. Carroll's group at Wake Forest University's Nano and Quantum Technologies Lab. It licenses IP directly related to this work from WFU. DL, SM, and GM are guest scientists at NQTL and were responsible for some spiral synthesis and MFM data collection. The manuscript and data collected from these studies were reviewed independently before submission.

**Acknowledgements**
The authors acknowledge support from Wake Forest University through the Nano and Quantum Technologies Laboratory as well as Quoherent Inc. We also thank Jordan Hatchel at Oak Ridge National Laboratory for assistance with the EELS and HRTEM imaging which was supported under the ORNL Center for Nanophase Materials Sciences, proposal CNMS2025-R-03097, and Chris Winkler at NC State for assistance with HRTEM work at the Analytical Instrumentation Facility (AIF) at North Carolina State University, which is supported by the State of North Carolina and the National Science Foundation (Award No. ECCS-2025064). The authors acknowledge student exchange with Frederich Alexander University in Erlangen, Germany allowing TC to spend three months with Prof. Krstic and Dr. Maria Gray.

## References

---

[1] Zhang, K., Lepage, H. V., Dong, Y. & Barnes, C. H. W. Charge qubits based on ultra-thin topological insulator films. Frontiers of Physics 19, (2024), Gerson J. Ferreira and Daniel Loss, Magnetically Defined Qubits on 3D Topological Insulators Phys. Rev. Lett. 111, 106802 (2013) DOI: https://doi.org/10.1103/PhysRevLett.111.106802, Hari P. Paudel and Michael N. Leuenberger, Three-dimensional topological insulator quantum dot for optically controlled quantum memory and quantum computing, Phys. Rev. B 88, 085316 (2013) DOI: https://doi.org/10.1103/PhysRevB.88.085316

[2] Liwei Jing, Mohammad Amini, Adolfo O. Fumega, Orlando J. Silveira, Jose L. Lado, Peter Liljeroth and Shawulienu Kezilebieke , Strain-induced two-dimensional topological crystalline insulator in bilayer SnTe, Nature Communications volume 17, Article number: 817 (2026) and Breunig, O., Ando, Y. Opportunities in topological insulator devices. Nat Rev Phys 4, 184–193 (2022). https://doi.org/10.1038/s42254-021-00402-6

[3] Timothy W. Carlson, Swathi Kadaba, Motahhare Mirhosseini, Maria Kolesnik-Gray, Gabriel Marcus, Lindsey J. Gray, Anthony Walsh, Vojislav Krstic, D. L. Carroll, Magnetic Force Imaging of 2D Topological Insulators, https://doi.org/10.48550/arXiv.2602.18281

[4] P. Michetti and P. Recher, Bound states and persistent currents in topological insulator rings, Phys. Rev. B 83, 125420 (2011) and Dun, C. et al. Bi2Te3 Plates with Single Nanopore: The Formation of Surface Defects and Self-Repair Growth. Chem. Mater. 30, 1965–1970 (2018).(for synthesis)

[5] Motahhare Mirhosseini, Swathi Kadaba, Allison Swyt, David L. Carroll, Geometry-tunable magnetic edge contrast in Bi2Te3 Corbino nanoplates, https://arxiv.org/pdf/2603.23693

[6] Thonikkadavan, A., d'Aquino, M. and Hertel, R. Rotating spin wave modes in nanoscale Möbius strips. npj Spintronics 4, 4 (2026). https://doi.org/10.1038/s44306-025-00118-8

[7] C Dun, CA Hewitt, Q Li, J Xu, DC Schall, H Lee, Q Jiang, DL Carroll, 2D chalcogenide nanoplate assemblies for thermoelectric applications, Advanced materials 29 (21), 1700070

[8] Li, C. & van 't Erve, O. & Li, Y. & Li, Lihaixiong & Jonker, Berend. (2016). Electrical Detection of the Helical Spin Texture in a p-type Topological Insulator Sb2Te3. Scientific Reports. 6. 29533. 10.1038/srep29533.

[9] Wang, G., Zhu, X., Wen, J. et al. Atomically smooth ultrathin films of topological insulator Sb2Te3 . Nano Res. 3, 874–880 (2010). https://doi.org/10.1007/s12274-010-0060-2

[10] Zhao Y, Zhang C, Kohler DD, Scheeler JM, Wright JC, Voyles PM, Jin S. Supertwisted spirals of layered materials enabled by growth on non-Euclidean surfaces. Science. 2020 Oct 23;370(6515):442-445. doi: 10.1126/science.abc4284. PMID: 33093106.

[11] Link, R., Marcus, G. & Carroll, D. Solvothermal growth of moiré superlattices in antimony telluride spiral-type nanoplates. Frontiers in Materials 9, (2022).

[12] See for instance; Asbóth, Oroszlány, and Pályi, A Short Course on Topological Insulators Lecture Notes in Physics 919 by Springer (Switzerland) 2016 ISBN 978-3-319-25605-4. There are also numerous ARPES studies cited in that text.

[13] Yosuke Kubota, Matthias Ludewig, Guo Chuan Thiang, Delocalized Spectra of Landau Operators on Helical Surfaces, Commun. Math. Phys. 395, 1211–1242 (2022) (DOI) https://doi.org/10.1007/s00220-022-04452-4

[14] Martin Lewin, Lars Mester, Tobias Saltzmann, Seung-Jae Chong, Marvin Kaminski, Benedikt Hauer, Marc Pohlmann, Antonio M Mio, Matti Wirtssohn, Peter Jost, Matthias Wuttig, Ulrich Simon, Thomas Taubner, Sb2Te3 Growth Study Reveals That Formation of Nanoscale Charge Carrier Domains Is an Intrinsic Feature Relevant for Electronic Applications, ACS Applied Nano Materials 2018 1 (12), 6834-6842

[15] Moon-ki Choi, Harley T. Johnson, Mechanics of out-of-plane screw dislocation in a 2D material, Extreme Mechanics Letters 77 (2025) 102320

[16] See for instance the many images presented in references 7 and 9. Notice that specifically in 9 where Moiré fringes are presented, the existence of dislocations within the crystal are quite clear.

[17] See for example: Topological Quantum Materials: Concepts, Models, and Phenomena 1st Edition by Grigory Tkachov Jenny Stanford Publishing, (2022) ISBN-13 : 978-9814968126

[18] Maissam Barkeshli and Xiao-Liang Qi, Topological Nematic States and Non-Abelian Lattice Dislocations, Phys. Rev. X 2, 031013 – (2012), DOI: https://doi.org/10.1103/PhysRevX.2.031013

[19] Ran, Y., Zhang, Y. & Vishwanath, A. One-dimensional topologically protected modes in topological insulators with lattice dislocations. Nature Phys 5, 298–303 (2009). https://doi.org/10.1038/nphys1220
[20] Erika N. Lima Tome M. Schmidt, and Ricardo W. Nunes, Topologically Protected Metallic States Induced by a One-Dimensional Extended Defect in the Bulk of a 2D Topological Insulator, Nano Lett. 2016, 16, 4025−4031DOI: 10.1021/acs.nanolett.6b00521
[21] Bin Xu, Sebastian Meyer, Matthieu J. Verstraete, Laurent Bellaiche, and Bertrand Dupé. First-principles study of spin spirals in the multiferroic BiFeO3, Physical Review B 103, 214423 (2021)
[22] Geometry-tunable magnetic edge contrast in Bi2Te3 Corbino nanoplates, M Mirhosseini, S Kadaba, A Swyt, DL Carroll arXiv preprint arXiv:2603.23693, and Timothy W Carlson, Motahhare Mirhosseini, Swathi Kadaba, Maria Kolesnik-Gray, Gabriel Marcus, Lindsey J Gray, Anthony Walsh, Vojislav Krstic, DL Carroll Magnetic Force Imaging of 2D Topological Insulators, arXiv preprint arXiv:2602.18281.
[23] Serra A. Electromagnetic Induction and the Conservation of Momentum in the Spiral Paradox. arXiv preprint physics/0012009. 2000 Dec 5.
[24] This follows reference [22] and is based on the methodology of Fujio Wakaya, Kenta Oosawa, Masahiro Kajiwara, Satoshi Abo, Mikio Takai, Detection of nonmagnetic metal thin film using magnetic force microscopy, Appl. Phys. Lett. 113, 261601 (2018
[25] T. Kobayashi, H. Mine, T. Tokuda, Y. Hashimoto, S. Katsumoto, and J. Haruyama, Edge-derived magnetisms in very thin non-doped Bi2Te3 nanomesh, Appl. Phys. Lett. 115, 093101 (2019); doi: 10.1063/1.5100024
[26] Guo-Hui Dong, Ying-Jie Zhu and Li-Dong Chen, Microwave-assisted rapid synthesis of Sb2Te3nanosheets and thermoelectric properties of bulk samples prepared by spark plasma sintering, J. Mater. Chem., 2010,20, 1976-1981 https://doi.org/10.1039/B915107A
[27] Zhen Li, Naihua Miao, Jian Zhou, Zhimei Sun, Zikui Liu, Huibin Xu, High thermoelectric performance of few-quintuple Sb2Te3 nanofilms, Nano Energy 43 (2018) 285–290 https://doi.org/10.1016/j.nanoen.2017.11.043
[28] Siegmar Roth and David Carroll, One-dimensional Metals 2nd edition (2006) Springer (Switzerland).
[29] While for a 3D material the formation of a Schottky barrier and subsequent rectification is well known, in low dimensional systems, the effects of electron transfer onto the object is more subtle. This was well explored for carbon nanotube systems in the early days of transport measurements in those systems. A terrific, modern exploration of this is found in the reference: Hu, S., Luo, K., Zhao, H. et al. Charge transport in carbon nanotube field-effect transistors: foundations for carbon-based nanoelectronics. Moore More 3, 2 (2026). https://doi.org/10.1007/s44275-026-00042-x
[30] Shivaji M. Sonawane, Shweta Chaure, Nandu B. Chaure, Characterization of Sb2Te3 thin films prepared by electrochemical technique, Journal of Physics and Chemistry of Solids, Volume 172, 2023, 111095, ISSN 0022-3697, https://doi.org/10.1016/j.jpcs.2022.111095.
[31] LibreTexts Chemistry B1: Workfunction Values (Reference Table): https://chem.libretexts.org/Ancillary_Materials/Reference/Reference_Tables/Bulk_Properties/B1%3A_Workfunction_Values_(Reference_Table)
[32] Rao F, Song Z, Cheng Y, Liu X, Xia M, Li W, Ding K, Feng X, Zhu M, Feng S. Direct observation of titanium-centered octahedra in titanium-antimony-tellurium phase-change material. Nat Commun. 2015 Nov 27;6:10040. doi: 10.1038/ncomms10040. PMID: 26610374; PMCID: PMC4674681.
[33] H. B. Zhang, H. L. Yu andG. W. Yang 2011 EPL 95 56002, doi:10.1209/0295-5075/95/56002
[34] W. S. Kim, G. Y. Chao, Louis J. Cabri, Phase relations in the Pd-Te system August 1990 Journal of the Less Common Metals 162(1):61-74, DOI:10.1016/0022-5088(90)90459-W
[35] He, Q., Lai, Y., Lu, Y. et al. Surface Reactivity Enhancement on a Pd/$Bi_2Te_3$ Heterostructure through Robust Topological Surface States. Sci Rep 3, 2497 (2013). https://doi.org/10.1038/srep02497
[36] Hanwen Wang, Wenjun Cui, Weixiao Lin, Junjie Yang, Wen Zhao, and Xiahan Sang Theoretical and Experimental Investigation on 3d Transition Metal Anisotropic Diffusion in van Der Waals Layered Sb2Te3 The Journal of Physical Chemistry C 2024 128 (16), 6859-6867 DOI: 10.1021/acs.jpcc.4c00145
[37] Lan, Yucheng, Wang, Dezhi, Chen, Gaoke, Ren, Z. (2008). Diffusion of nickel and tin in p-type $(Bi,Sb)_2Te_3$ and n-type $Bi_2(Te,Se)_3$ thermoelectric materials. Appl. Phys. Lett.. 92. 101910.
[38] Satilmis Budak, Zhigang Xiao, Jorden Cole, Dennis Price, Tyler Davis, Tiara Strong, Mohammad A. Alim; Thermoelectric and optical properties of advanced thermoelectric devices from Ni/Bi2Te3/Ni and Ni/Sb2Te3/Ni thin films. J. Vac. Sci. Technol. B 1 September 2017; 35 (5): 051401. https://doi.org/10.1116/1.5001046
[39] Zhen-Wei Sun, Kai-Wen Cheng, Si-Wei Lin, V. K. Ranganayakulu, Yang-Yuan Chen, Shang-Jui Chiu, Tai-Wei Lee, and Albert T. Wu, Stoichiometric Effect of Sb2Te3 Thin Film on Thermoelectric Property ACS Appl. Energy Mater. 2022, 5, 7026−7033 https://doi.org/10.1021/acsaem.2c00657

[40] Bohan Zhang, Wei Zhu, Lili Cao, Yuedong Yu, Dongli Qin, Xin Huang, Yuan Deng, Toward Reduced Interface Contact Resistance: Controllable Surface Energy of Sb2Te3 Films via Tuning the Crystallization and Orientation ACS Applied Materials & Interfaces Vol 14 / Issue 8 February 15, 2022, https://doi.org/10.1021/acsami.1c22908

[41] See for example:

a. Liu, C.-X. et al. Oscillatory crossover from two-dimensional to three-dimensional topological insulators. Phys. Rev. B 81, (2010).
b. Hasan, M. Z. & Moore, J. E. Three-Dimensional Topological Insulators. Annu. Rev. Condens. Matter Phys. 2, 55–78 (2011).
c. Hasan, M. Z. & Kane, C. L. Colloquium: Topological Insulators. Rev. Mod. Phys. 82, (2010).
d. Ando, Y. Topological Insulator Materials. J. Phys. Soc. Jpn. 82, 102001 (2013).

[42] For example: Judy J. Cha, Kristie J. Koski, Kevin C. Y. Huang, Ken Xingze Wang, Weidong Luo, Desheng Kong, Zongfu Yu, Shanhui Fan, Mark L. Brongersma, and Yi Cui Two-Dimensional Chalcogenide Nanoplates as Tunable Metamaterials via Chemical Intercalation Nano Letters 2013 13 (12), 5913-5918 DOI: 10.1021/nl402937g (an example that helped guide our studies)

[43] Hachtel, J. A., Lupini, A. R. & Idrobo, J. C. Exploring the capabilities of monochromated electron energy loss spectroscopy in the infrared regime. Sci Rep 8, 5637 (2018).

[44] I. A. Nechaev, I. Aguilera, V. De Renzi, A. di Bona, A. Lodi Rizzini, A. M. Mio, G. Nicotra, A. Politano, S. Scalese, Z. S. Aliev, M. B. Babanly, C. Friedrich, S. Blügel, and E. V. Chulkov, Quasiparticle spectrum and plasmonic excitations in the topological insulator $Sb_2Te_3$, Phys. Rev. B 91, 245123 – Published 11 June, 2015 DOI: https://doi.org/10.1103/PhysRevB.91.245123

[45] Parivash Moradifar, Austin G. Nixon, Tiva Sharifi, Tim Brandt van Driel, Pulickel Ajayan, David J. Masiello, and Nasim Alem, Nanoscale Mapping and Defect-Assisted Manipulation of Surface Plasmon Resonances in 2D Bi2Te3/Sb2Te3 In-Plane Heterostructures, Adv. Optical Mater. 2022, 10, 2101968, DOI: 10.1002/adom.202101968

[46] Parivash Moradifar, Plasmonic Resonances In 2D and 3D Nanostructures Investigated By Monochromated Electron Energy Loss Spectroscopy (Mono-EELS), PhD. Dissertation in Materials Science and Engineering, Pennsylvania State University (2020)

[47] Henning Schlömer, Zhihao Jiang, and Stephan Haas, Plasmons in two-dimensional topological insulators, Physical Review B 103, 115116 (2021) DOI: 10.1103/PhysRevB.103.115116

[48] Lai, Yi-Ping & Lin, I-Tan & Wu, Kh & Liu, Jiaming. Plasmonics in Topological Insulators. Nanomaterials and Nanotechnology. 4. 1. (2014). DOI 10.5772/58558.

[49] For example: Hatem Diaf, Antonio Pereira, Patrice Melinon, Nicholas Blanchard, Florent Bourquard, Florence Garrelie, Christophe Donnet, and Martin Vondráčk, Revisiting thin film of glassy carbon Phys. Rev. Materials 4, 066002 – Published 5 June, 2020, DOI: https://doi.org/10.1103/PhysRevMaterials.4.066002

[50] E.H. Alizade, Study of plasmon resonance in Bi2Se3 and Sb2Te3 by infrared spectral ellipsometry, Optics and Spectroscopy, 2022, Vol. 130, No. 2

[51] Cana Elgvin, Fredrik S. Hage, Øystein Prytz, Khairi Elyas, Katja Höflich, Christoph T. Koch and Hannah C. Nerl, Advances in momentum-resolved EELS of phonons, excitons and plasmons in 2D materials and their heterostructures, Phys. Rev. Materials 10, 020201 – Published 4 February, 2026 DOI: https://doi.org/10.1103/8wwb-lxk2 and O.L. Krivanek, N. Dellby, T.C. Lovejoy, N.J. Bacon, G.J. Corbin, P. Hrncirik, Z.S. Szilagyi, T. Aoki, R.W. Carpenter, P.A. Crozier, J. Zhu, P. Rez, R.F. Egerton, and P.E. Batson, Exploring Phonon Signals by High Energy / High Spatial Resolution EELS, Microsc. Microanal. 20 (Suppl 3), 2014 doi:10.1017/S1431927614002050

[52] Katsura Norimatsu; Masaki Hada; Shuhei Yamamoto; Takao Sasagawa; Masahiro Kitajima; Yosuke Kayanuma; Kazutaka G. Nakamura, Dynamics of all the Raman-active coherent phonons in Sb2Te3 revealed via transient reflectivity, J. Appl. Phys. 117, 143102 (2015) https://doi.org/10.1063/1.4917384

[53] J.Mertens, P.Kerres, Y.Xu, M.Raghuwanshi, D.Kim, C.-F.Schön, J.Frank, F.Hoff, Y.Zhou, R.Mazzarello, A. R.Jalil, M.Wuttig, Confinement-Induced Phonon Softening and Hardening in Sb2Te3 Thin Films. Adv. Funct. Mater.2024, 34, 2307681. https://doi.org/10.1002/adfm.202307681

[54] Maisel Licerán L, García Flórez F, Siebbeles LDA, Stoof HTC. Single-particle properties of topological Wannier excitons in bismuth chalcogenide nanosheets. Sci Rep. 2023 Apr 18;13(1):6337. doi: 10.1038/s41598-023-32740-z. PMID: 37072513; PMCID: PMC10113227.

[55] Reutzel, M., Jansen, G. S. M., & Mathias, S. (2024). Probing excitons with time-resolved momentum microscopy. Advances in Physics: X, 9(1). https://doi.org/10.1080/23746149.2024.2378722

[56] Hao Chen, Zhou Cui, Rui Xiong, Huiwen Yu, Jiaxiang Li, Jiansen Wen, Xiaochun Huang, Cuilian Wen and Baisheng Sa, Two-dimensional Si2Te2 monolayers and Si2Te2/

Sb2Te3 van der Waals heterostructures: promising infrared photodetector materials, J. Mater. Chem. C, 2025, 13, 15372 DOI: 10.1039/d5tc01609f

[57] Lin Gu, Vesna Srot, Wilfried Sigle, Christoph Koch, Peter van Aken, Ferdinand Scholz, Sarad B. Thapa, Christoph Kirchner, Michael Jetter, and Manfred Rühle, Band-gap measurements of direct and indirect semiconductors using monochromated electrons, Physical Review B 75, 195214 2007 DOI: 10.1103/PhysRevB.75.195214

[58] B. A. Bernevig, T. L. Hughes and S.-C. Zhang, "Quantum spin Hall effect and topological phase transition in HgTe quantum wells," Science, vol. 314, p. 1757–1761, 2006.

[59] A. Michetti and P. Recher, "Corbino disk in quantum spin Hall insulators," Physical Review B, vol. 83, p. 125420, 2011.